\documentclass[11pt,a4paper]{article}
\usepackage{cite}
\usepackage{amsmath}
\usepackage{amssymb}
\usepackage{pstricks}
\newcommand{\ii}{\mathrm{i}}
\newcommand{\na}{\nabla}
\newcommand{\dd}{\mathrm{d}}
\newcommand{\pd}{\partial}
\newcommand{\hh}{\mathcal{H}}
\newcommand{\F}{\mathcal{F}}
\newcommand{\M}{\mathcal{M}}

\newcommand{\e}{\mathrm{e}}
\newcommand{\ket}[1]{\left|#1\right\rangle}
\newcommand{\bra}[1]{\left\langle #1\right|}
\newcommand{\bracket}[2]{\left\langle%
#1\left.\right|#2\right\rangle}
\newcommand{\tr}{\mathop{\mathrm{tr}}}

\newcommand{\const}{\mathrm{constant}}
\newcommand{\A}{\mathfrak{A}}
\newcommand{\B}{\mathfrak{B}}
\newcommand{\I}{\mathbb{I}}
\newcommand{\R}{\mathbb{R}_\theta^n}
\newcommand{\x}{\mathbf{x}}
\newcommand{\p}{\mathbf{p}}
\newcommand{\q}{\mathbf{q}}
\renewcommand{\a}{\mathbf{a}}
\newcommand{\N}{\mathbf{N}}
\newcommand{\su}{\mathrm{su}}
\newcommand{\U}{\mathrm{U}}
\newcommand{\SO}{\mathrm{SO}}
\newcommand{\dop}{\hat{\delta}}
\newcommand{\spec}{\mathop{\mathrm{spec}}}
\newcommand{\pr}[1]{\langle #1 \rangle_{\A}}
\newcommand{\prB}[1]{\langle #1 \rangle_{\B}}

\newcommand{\End}{\mathop{\mathrm{End}}_{\A}}
\newcommand{\ov}[2]{\genfrac{}{}{0pt}{1}{#1}{#2}}
\newtheorem{exercise}{Exercise}
\numberwithin{equation}{section}

\begin{document}
\title{Gauge Invariance and Noncommutativity\thanks{Work supported by
RFBR grant \#, INTAS grant \# and Scientific School Support grant
\#.}}


\author{\textbf{Corneliu Sochichiu}\\
Institutul de Fizic\u a Aplicat\u a A\c S,\\ str. Academiei, nr.
5, Chi\c sin\u au, MD2028\\
MOLDOVA\\
Bogoliubov Laboratory of Theoretical Physics\\
Joint Institute for Nuclear Research\\ 141980 Dubna, Moscow Reg.\\
RUSSIA\\
email: \texttt{sochichi@thsun1.jinr.ru}%
}
\maketitle
\begin{abstract}
The role of the gauge invariance in noncommutative field theory is
discussed. A basic introduction to noncommutative geometry and
noncommutative field theory is given. Background invariant
formulation of Wilson lines is proposed. Duality symmetries
relating various noncommutative gauge models are being discussed.
\end{abstract}
\newpage
\tableofcontents

\newpage
\section{Introduction}

Understanding the structure of the space-time has always been a
challenge for the theoretical physics.

Since Einstein the gravity interactions are understood as
deformation of the space-time metric. The description in the
framework of \emph{General Relativity\/} gives satisfactory
results for a wide range of scales: from Plank to that Galaxy
sizes and more which covers, in fact, all experimentally and
astrophysically accessible scales up to date. However, the sizes
smaller than the Plank scale remain beyond the consistency limit
of General Relativity. Due to a number of problems the General
Relativity appeared to be incompatible with the quantum approach
thus making such extension impossible.

The lack of reliable data on the space-time structure at very
small sizes (or very high energies) provides a good ground for
various models, even extremely exotic ones, to be constructed. In
this context the geometry including the topology can be accepted
to have a form which is quite different from what we are
``familiar'' with at larger scales. In this case the ``familiar''
geometry is obtained as an ``average'' of such exotic structures.

As far as in 50's of the last century it was proposed to consider
the possibility that at small distances coordinates satisfy the
Heisenberg-like commutation relation rather than commute ``as
usually'' \cite{Snyder:1947a,Snyder:1947b}. Surprisingly, it
appeared that such ``spaces'' possess many features allowing to
build over them quantum field theory as it were an ordinary space.
Even more surprising is that these models revived recently,
however, from different considerations.

An alternative approach to get a quantum description of gravity,
or the description of the space-time structure at very small
distances is to consider the gravity as an effective model to one
which is a more ``quantum-friendly''. Such a model was discovered
some 30 years ago and it is known as \emph{String Theory\/} or its
supersymmetric extension(s): \emph{Superstring Theory}.

This conceptually simple model of a relativistic one-dimensional
extended object moving in a $D$-dimensional space-time (target
space) appears to have extremely reach and complicate structure.
The self-consistence of the quantum model impose tight conditions
on the geometry of the target space. Thus, cancellation of the
conformal (modular) anomaly requires the space-time dimensionality
to be 26 in the bosonic case or 10 in the supersymmetric one.

The mathematical rigidity of the string theory allows the only
bosonic string model (which, however, appears to be unstable due
to the presence of a tachyonic mode in the spectrum), and a few
supersymmetric ones, basically they are, type I, type IIA, type
IIB, heterotic string and their modifications. These models differ
by the number and relative chirality of supercharges, but they are
related with each other by \emph{duality\/} relations. Basing on
this duality symmetry it was conjectured that all above string
models (maybe, except the bosonic string) are different
perturbative regimes of the same larger theory called
\emph{M-theory}, about which we know that its low energy effective
field theory is the eleven dimensional supergravity.

One conceptual problem arising when the superstring models are
proposed as fundamental theory is that they are multi dimensional
while the observable world has only four dimensions. This problem
was proposed to be solved by compactification extra dimensions to
sizes less than observable by our today's tools. Although this
mechanism could provide an explanation of the four-dimensionality
of the low energy world, but until now there is no clear mechanism
demonstrating why and how this realizes dynamically in the string
theory.

Developing of the string theory in recent years showed that the
compactification is not the only way to obtain the
four-dimensional world from the string theory. Nonperturbative
analysis of the string dynamics unveiled a large set of extended
objects of various dimension in the nonperturbative spectrum of
the string theory. These objects are called D-branes as it comes
from [mem]brane. In particular, existence of D-branes is required
by duality symmetries \cite{Polchinski:1995mt,Polchinski:1996na}
(see also \cite{Polchinski:1998rr,Polchinski:1998rq}.

It was also advocated
\cite{Seiberg:1999vs,Connes:1998cr,Ho:1997jr,Ho:1998yk,Chu:1998qz}
that in the background of constant stringy NSNS field $B_{\mu\nu}$
the low energy dynamics of the nonperturbative objects is
described by field theory models on noncommutative spaces. Namely,
the quantum algebra of string observables is such that in the
limit of large constant $B_{\mu\nu}$ the subalgebra corresponding
to the dynamics of the string ends decouples and can be described
by a field theory on a space whose coordinates satisfy,
\begin{equation}\label{comm1}
  [x^\mu,x^\nu]=\ii \theta^{\mu\nu},
\end{equation}
where $\theta^{\mu\nu}\neq 0$ is a constant matrix depending on
$B_{\mu\nu}$. In such a way we ``rediscovered'' the noncommutative
space-time as a consequence of the string description of gravity.

It appears that in many cases it is possible to develop an
analysis of field theoretical models over noncommutative space
very similar to one we have in ordinary spaces. This analysis
sometimes unveils very surprising properties of such models. Thus,
even in classical or tree level these models are very different
from their relatives on commutative spaces. Among such properties
one could enumerate the so called IR/UV mixing, which consists in
dependence of the small distance behaviour on large distance data,
noncommutative solitons and lump configurations, which are
localized solutions of the size of noncommutativity and duality
relations which we are considering in the present paper. Here we
discuss the noncommutative theories in the classical approach. In
perturbation theory there are additional problems which we are not
going to discuss here, the relevant reference is
\cite{Minwalla:1999px,VanRaamsdonk:2000rr} and consecutive works.

The plan of the paper is the following. The first part is an
introductory one. There we introduce the notion of noncommutative
space, including a brief review of the Connes concept of the
noncommutative space. As an illustration we consider a lattice
example. We also consider the simplest examples of noncommutative
spaces such as noncommutative plane $\R$, noncommutative torus
$\mathbb{T}^n_q$, and noncommutative or fuzzy sphere
$\mathbb{S}^n_\theta$. The last two noncommutative spaces can be
embedded into the first one. We review the notion of Weyl
ordering, Weyl symbols and introduce noncommutative calculus. This
allows one the introduction of the notion of noncommutative field
theory.

The second part is devoted exclusively to the field theory. Here
we introduce scalar field. Analyzing it we conclude that it
possess a broken gauge invariance which is restored by
introduction of the gauge field. In the noncommutative theory the
gauge symmetry plays the role of representation independence and
is fundamental to the model. (The algebra \eqref{comm1} allows a
class of unitary equivalent representations, while for $\theta=0$
one has the only representation.) Respectively the role of gauge
fields appears to be extremely important in noncommutative
theories. Thus, as we will see different gauge field
configurations can be interpreted as different noncommutative
spaces. A particular interest present the constant curvature ones
which generate flat noncommutative spaces or ``vector bundles''
over flat noncommutative spaces. This relates, in fact, gauge
models on different noncommutative spaces having different
noncommutative structure, different local gauge group or even
different dimensionality. This equivalence is similar to duality
relations in string theory. According to it many different
noncommutative gauge models appear to be different perturbative
limits of the same model. This model appears to be the infinite
dimensional Hilbert space version of the IKKT or BFSS matrix model
\cite{Ishibashi:1996xs,Banks:1997vh} depending if there is a
commutative time or not.

As this paper was initially intended as lecture notes, we supply
each section with exercises.

Another reviews on noncommutative field theory can be found in
\cite{Harvey:2001pd,Harvey:2001yn,Konechny:2000dp,Konechny:2001wz,%
Nekrasov:2000ih,Szabo:2001kg,Lizzi:2001nd,Arefeva:2001ps}.

\newpage
\section{Noncommutative Space}\label{section:nc-space}
\subsection{The Connes' Concept of the Noncommutative Space}

In this section we give a brief description of Connes idea of
noncommutative geometry. For details and mathematical rigor
readers are referred to the original Connes book
\cite{Connes:book} as well as to later review papers e.g.
\cite{Connes:2000by}. Since in the remaining part of this paper we
will work with simplest examples and will not make use of  of
noncommutative spaces this section has mainly a ``philosophical''
character and therefore may be skipped without injury the
understanding the rest. However, we decided to put it here, since
the Connes formalism gives a natural way  to generalize the
analysis developed in next sections (but not necessarily an easy
one).

Let us start the applying Connes description first to a
commutative Riemann or spin manifold.

It is considered that the Connes approach comes from a statement
that \emph{a topological space can be recovered from the algebra
of continuous functions on it.} (Which is a topological space
too!) In other words, knowing all possible ways a topological
space can be continuously mapped into $\R{}$ or $\mathbb{C}$ we
know the topological space itself.

This idea seems plausible since the algebra of functions is much
``bigger'' than the space itself, but let illustrate it on the
following discrete example.

\begin{figure}[t]
 \centering{\input{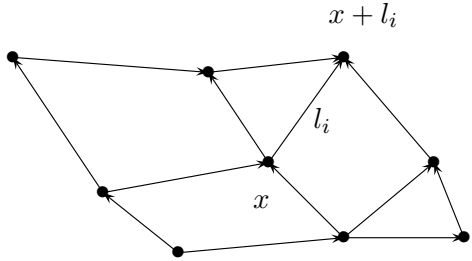}}
\caption{An example of irregular lattice. Its topology is given by
the neighborhood relations of points. Neighbor points are those
connected by links. In order to count all links just once, they
are given the orientation denoted by arrows. The lengths of links
introduce the metric on the lattice.}\label{Fig:lattice}
\end{figure}

Consider a (possibly irregular) lattice $\Gamma$ like in
Fig.\ref{Fig:lattice}.

The topology on the lattice is given by characteristic of two
points as being neighbor or not, which graphically can be
represented by drawing a link connecting two neighbor points.
Thus, the lattice topology describes how points are arranged with
respect to each other. Obviously, this arrangement is invariant
against small changes in the positions of points.

Lattice analog of continuous function is a map from $\Gamma$ to
$\mathbb{Z}$ which translates neighbor points of $\Gamma$ to
neighbor points of $\mathbb{Z}$, which in the natural topology of
$\mathbb{Z}$ are $n$ and $n+1$ for some $n\in\mathbb{Z}$. All such
functions for an algebra $\A$ with respect to multiplication and
sum. The induced topology on the algebra of such functions is
described as follows. Two maps are neighbor on a point if they map
it to mutually neighbor points.

Let us note that all maps that send given point $x\in\Gamma$ to
the origin $0\mathbb{Z}$ (vanish on $x$) form an ideal $\A_x$ of
the above algebra. Conversely, to each minimal ideal $\A_x$ of
algebra of functions one can put into correspondence a point
labeling it. (Alternatively, one could consider ideals of
functions vanishing everywhere except $x$, but in this case these
are represented by functions in continuous case.) Thus, counting
all minimal ideals of algebra $\A$ give us the lattice $\Gamma$ as
a simple set of points.

The topology of the set of $\Gamma$ is recovered as follows. Two
points $x$ and $y$ represented by $\A_x$ and $\A_y$ iff any two
elements $f\in\A_x$ and $g\in\A_y$ are neighbor on $\Gamma$.

The continuous picture one can obtain by assuming continuum limit
of the above, e.g. when the lattice points become infinitely close
to each other.

The next question which may arise is if it is possible to restore
more than that? The answer to this is positive. For example one
can recover the metric if an additional structure to the algebra
of continuous functions is assumed. Indeed, let us assume that
each point has a fixed number $l$ of neighbors and introduce a set
of operators $\ii p_i$ $i=1,\dots, l$, which substitute the point
$x$ by its $i$-th neighbor: $x\to x+l_i$ and divides each function
on $l_i$:
\begin{equation}\label{p-mu}
  [\ii p_i, f](x)=\frac{1}{l_i}(f(x+l_i)-f(x)),
\end{equation}
where $l_i$ is the distance between two neighbors $x$ and $x+l_i$.

Having such operators in addition to the algebra of functions $\A$
we can extract the information about the distance on the lattice.
Indeed, equation,
\begin{equation}\label{distance}
  d(x,y)=\sup_{|\sum_i[p_i,f]^2|\leq 1} |f(x)-f(y)|,
\end{equation}
in the case of neighbor points gives the correct answer --
$d(x,x+l_i)=l_i$.

To obtain the continuum limit one has to send $l_i\to 0$ in a way
that ensures that lines connecting neighbor points form continuous
paths which become coordinate lines. In this case, the commutators
$[\ii p_i,f]$ become derivatives with respect to this coordinate
lines. Physically the continuum limit is reached when all $l_i$
become smaller than the typical physical scale $l_{\mathrm{ph}}$.
In the same manner one can encode as many structures of the
manifold as one wishes by adding respective operator to the
algebra of functions, e.g. the spin structure can be recovered
from the Dirac operator.

In example we considered the algebra $\A$ of functions is a
commutative algebra. The next step in the generalization is to
consider an arbitrary noncommutative algebra instead. For
noncommutative algebra an additional input is needed its
representation. (For commutative algebra the representation was
trivial.)

Thus, we came to the Connes definition of the noncommutative space
or the Connes triple, which is
\begin{equation}\label{triple}
  \M=(\A, D, \hh),
\end{equation}
where $\A$ is an algebra of bounded operators (irreducibly)
represented on the Hilbert space $\hh$, which is not necessarily
finite-dimensional\footnote{We will consider the cases where it is
infinite-dimensional separable.} and $D$ is an operator (not
necessarily bounded) acting self-adjointly on $\hh$. $\A$ plays
the role of the algebra of functions over noncommutative space
while $D$ one of a differential operator e.g. for a spin manifold
it is Dirac operator.

In the next section we will consider the simplest cases of flat
noncommutative spaces, but one should keep in mind that this space
is given by the triple \eqref{triple}, and therefore changing a
component of \eqref{triple} results in change of the
noncommutative space.

\subsection{Simplest Cases of Noncommutative Spaces}

As usual happens from the whole variety of mathematical tools only
simplest ones are used in most physical applications. In the
present development of the noncommutative field theory only
noncommutative plane $\R$, noncommutative torus $T_\theta^n$ and
noncommutative, or fuzzy sphere are most commonly used. Therefore,
we describe this cases in more details.

\subsubsection{Noncommutative $\R$}
Let us start with noncommutative $\R$. Consider the algebra
generated by $\x^\mu$, $\mu=1,\dots,p$, subject to commutation
relations,
\begin{equation}\label{Heis1}
  [\x^\mu,\x^\nu]=\ii\theta^{\mu\nu},
\end{equation}
where $\theta^{\mu\nu}$ are commutative elements (c-numbers)
forming a nondegenerate (with respect to indices $\mu$ and $\nu$)
matrix. As a consequence we have that $p$ should be always even.

Generators $\x^\mu$ are called noncommutative coordinates of $\R$,
while $\theta^{\mu\nu}$ is the noncommutativity matrix. In the
case of degenerate noncommutativity matrix one can pass to
generators $x^i$ and $\x^\alpha$, $\alpha=1,\dots,p'\leq p$,
$i=p'+1,\dots,p$, where $x^i$ corresponding to zero modes of
$\theta^{\mu\nu}$ form a commutative subspace of $\R$, while
$\x^\alpha$ is its completion where the restriction of
$\theta^{\mu\nu}$ is nondegenerate.

Commutation relations \eqref{Heis1} correspond to the standard
$p/2$-dimensional Heisenberg algebra, although in a nonstandard
parameterizations. From the courses on Quantum Mechanics we are
familiar with the form where $\x^\mu$ is split into pairs of
$\p_i$ and $\q^i$, $i=1,\dots,p/2$, satisfying,
\begin{equation}\label{stand-Heis}
  [\p_i,\q^j]=-\ii \hbar \delta_i^j.
\end{equation}
In our case the rotations of the coordinates $x^\mu$ allow to
bring the algebra \eqref{Heis1} to the form with ``Plank
constant'' depending on direction,
\begin{equation}\label{Heis2}
  [\p_i,\q^j]=-\ii \theta_{(i)} \delta_i^j,
\end{equation}
we denoted it by $\theta_{(i)}$. Absorption of $\theta_{(i)}$ is
possible in principle by scaling $p$ and $q$, but this
transformation alters the metric of $\R$.

Thus, we got a familiar thing, Heisenberg algebra, whose
irreducible representations, we hope, are well known from standard
QM courses. They are all \emph{infinite dimensional\/} and
\emph{unitary equivalent\/} to each other.\label{unit-eq}

The irreducibility in particular mean that any operator commuting
with all $x^\mu$ is a $c$-number in the sense that it is
proportional to the unity operator. Later, this property will play
an important role in our analysis, as it means that any such
operator can be expressed as an operator function of $X^\mu$.

Consider now $\mathbb{R}^2_\theta$, or noncommutative plane as the
simplest example. It is given by ``coordinates'' $x^{1,2}$
satisfying the algebra,
\begin{equation}\label{12}
  [\x^1,\x^2]=\ii \theta,\qquad \theta\neq 0.
\end{equation}
This is usual Heisenberg algebra of one-dimensional quantum
mechanics with $\q=\x^1$, $\p=\x^2$ and the Planck constant equal
to $\theta$. The representation of algebra \eqref{12} is realized
on e.g. by $L^2$-integrable functions of $\q=\x^1$, on which
$\x^1$ acts by multiplication by its eigenvalue and $\x^2$ as
derivative,
\begin{align}
 &\x^1 f(x^1)=x^1 f(x^1),\qquad f\in L^2,\\
 &\x^2 f(x^1)=-\ii\theta\frac{\pd}{\pd x^1}f(x^1).
\end{align}

The analog of noncommutative complex coordinates is given by the
oscillator representation. It is given in terms of operators $\a$
and $\bar{\a}$ defined as follows,
\begin{equation}\label{aabar}
  \a=\sqrt{\frac{1}{2\theta}}(\x^1+\ii \x^2),\qquad
  \bar{\a}=\sqrt{\frac{1}{2\theta}}(\x^1-\ii \x^2).
\end{equation}
These operators satisfy the commutation relation,
\begin{equation}\label{[aa]}
  [\a,\bar{\a}]=1,
\end{equation}
and can be represented in terms of oscillator basis $\ket{n}$
given by the eigenvectors of operator $\N=\bar{\a}\a$, while $\a$
and $\bar{\a}$ act as lowering and rising operators,
\begin{equation}\label{osc-bas}
  \a\ket{n}=\sqrt{n}\ket{n-1},\qquad
  \bar{\a}\ket{n}=\sqrt{n+1}\ket{n+1}.
\end{equation}

In general case of $\R$ one can introduce ``complex'' coordinates
as well. The are given by multi-dimensional oscillator operators
$\a_i$ and $\bar{\a}_i$ defined by,
\begin{equation}\label{ai}
  \a_i=\sqrt{\frac{1}{2\theta_{(i)}}}(\q^i+\ii \p_i),\qquad
  \bar{\a}_i=\sqrt{\frac{1}{2\theta_{(i)}}}(\q^i-\ii \p_i),
\end{equation}
and satisfying the commutation relations,
\begin{equation}\label{[aai]}
  [\a_i,\bar{\a}_j]=\delta_{ij}.
\end{equation}
The natural basis is given by states $\ket{\vec{n}}$ which are
eigenvectors of operators $\vec{\N}$ with ``components''
$\N_i=\bar{\a}_i\a_i$ (no sum over $i$). Vector $\vec{n}$ takes
values on an infinite positive $p/2$ dimensional rectangular
lattice, $\vec{n}=\sum_i n_i\vec{e}_i$, $n_i\geq 0$. Operators
$\a_i$ and $\bar{\a}_i$ are respectively lowering and rising
operators for eigenvalue $n_i$,
\begin{equation}\label{p/2-osc}
  \a_i\ket{\vec{n}}=\sqrt{n_i}\ket{\vec{n}-\vec{e}_i},\qquad
  \bar{\a}_i\ket{\vec{n}}=\sqrt{n_i+1}\ket{\vec{n}+\vec{e}_i}.
\end{equation}

\subsubsection{Noncommutative $\mathbb{T}_\theta^n$}
In the commutative case one can pass from the plane to torus by
factorizing the plane with respect to a discrete group of the
translation group of the plane. In the simplest case this is
obtained by the identification
\begin{equation}\label{com-torus}
  x^\mu\sim x^\mu+l^\mu.
\end{equation}

A similar thing can be done also in the noncommutative case. The
identification like \eqref{com-torus} can be done if one considers
the noncommutative algebra of functions as being generated by
operators,
\begin{equation}\label{nc-torus1}
  U_\mu=\e^{2\pi\ii \frac{x^\mu}{l^\mu}},\qquad \mu=1,\dots,p,
\end{equation}
which are noncommutative ``coordinates'' of the torus.

The generators $U_\mu$ satisfy,
\begin{equation}\label{torus-alg}
  U_\mu U_\nu =q_{\mu\nu}U_\nu U_\mu,
\end{equation}
where,
\begin{equation}\label{q}
  q_{\mu\nu}=\e^{-\ii\frac{(2\pi)^2}{l^\mu l^\nu}\theta^{\mu\nu}}\equiv
  \e^{2\pi\ii \Xi^{\mu\nu}}\qquad \text{(no sum)}.
\end{equation}

The above equations were computed using the
Cambell--Bjorken--Haus\-dorf formula,
\begin{equation}\label{cbh}
  \e^A \e^B=\e^{A+B+\frac{1}{2}[A,B]+\dots},
\end{equation}
which is also a useful tool for our analysis.

In general one can define the noncommutative torus by the algebra
\eqref{torus-alg} for some $\Xi^{\mu\nu}$ with no direct reference
to the noncommutative plane. From the commutative quantum field
theory we are familiar that torus provides an IR regularization of
theories on the plane.

In the noncommutative case it turns out that the noncommutative
torus provide not only IR but also the UV regularization due to
the phenomenon known as \emph{IR/UV mixing\/}. Moreover, it
appears that in the special case when $\Xi^{\mu\nu}$ is a rational
valued matrix, the respective irreducible representation become
finite dimensional. This kind of noncommutative lattice theories
were also studied (see Refs.
\cite{Ambjorn:2000nb,Ambjorn:2000cs}).

From the other hand, the noncommutative torus can be always embedded
into the noncommutative plane of the same dimension. Due to this
property in the rest of this paper we will consider mainly the
theories on noncommutative planes.

\subsubsection{Noncommutative or Fuzzy $\mathbb{S}^n$}
Another natural example of compact noncommutative space is
noncommutative, or fuzzy sphere. Here we describe first the
noncommutative analog of the two dimensional sphere.
Noncommutative sphere of arbitrary dimensions is obtained in a
similar way.

Two dimensional commutative sphere can be defined in multiple
ways, for example like a factor $\SO(3)/\SO(2)$, or like a
submanifold of $\mathbb{R}^3$, defined by condition $x^2=r^2$.
Both these possibilities can be generalized to the noncommutative
case. In what follows choose the second one.

Consider noncommutative $\mathbb{R}^3$ with a nonconstant
noncommutativity,
\begin{equation}\label{nonconst-r3}
  [\x^i,\x^j]=\ii \alpha\epsilon^{ijk}\x^k,\qquad 1=1,2,3,
\end{equation}
where $a$ is a dimensional parameter. The commutation relations
\eqref{nonconst-r3} are nothing else than ones of $\su(2)$
algebra. Irreducible representations of algebra
\eqref{nonconst-r3} are possible when,
\begin{equation}\label{r2=a2}
  r^2=\alpha^2j(j+1),
\end{equation}
where $l$ can be half integer, which restricts the radius of
noncommutative sphere to quantized values in terms of $\alpha$.
(Or, equivalently, $\alpha$ is restricted in terms of $r$.)

Therefore, the the two dimensional noncommutative sphere of the
radius $r^2=\alpha^2j(j+1)$ is given by the algebra of operators
acting on the irreducible representation of $\su(2)$ of the spin
$j$.

Again, the data corresponding to the noncommutative sphere can be
expressed in terms of those of noncommutative plane. In terms of
oscillator operators one can write the $\su(2)$ algebra
\eqref{nonconst-r3} in the Chevalley basis $x_{\pm},x_3$ as
follows,
\begin{equation}\label{L+-}
   (x_\pm/\alpha)=\bar{a}\sqrt{\frac{j(j+1)-(\N-j)(\N-j\pm 1)}{\N}},\qquad
   (x_3/\alpha)=\N-j,
\end{equation}
where $\N$ is the oscillator number operator. Operators
\eqref{L+-} generate an irreducible representation on the first
$2j+1$ oscillator states.
\begin{exercise}
  Show this.
\end{exercise}

The $n$-dimensional noncommutative sphere is given by
$\SO(n+1)/SO(n)$. Therefore, one can define it trough irreducible
representations of $\SO(n+1)$ which are singlet in the factor
$\SO(n)$.
\begin{exercise}
  How many such representations are there?
\end{exercise}
\subsection{Weyl Ordering}
In the remaining part of this section we discuss noncommutative
planes.

So far, we defined the noncommutative space by means of its
noncommutative algebra of functions and its representation. When
working with functions in commutative space we prefer to operate
with explicit $x$-dependent form rather then with abstract algebra
element. As it turns the same possibility exists also in
noncommutative geometry. It is given by \emph{Weyl ordering\/} and
respective objects are \emph{Weyl symbols\/}. Roughly speaking the
Weyl symbol of an operator $\hat{\mathcal{O}}(\x)$ with respect to
the set of basic operators $\x^\mu$ is the function
$\mathcal{O}(x)$ symmetrized with respect to all $\x^\mu$. Since
in this forms functions do not depend on the commutation relations
$\x^\mu$ have one can substitute the operators by usual functions
by the following rule,
\begin{subequations}\label{Weyl-poly}
\begin{align}\label{x->x}
  &\x^\mu\to x^\mu, \\ \label{xx->xx}
  &\x^\mu\x^\nu=\frac{1}{2}(\x^\mu\x^\nu+\x^\nu\x^\mu)+
  \frac{\ii}{2}\theta^{\mu\nu}\to x^\mu x^\nu+
  \frac{\ii}{2}\theta^{\mu\nu},\\ \nonumber
  &     \dots
\end{align}
\end{subequations}
Here $x^\mu$ are treated as ordinary commutative coordinates.

The transformation \eqref{Weyl-poly} defines the Weyl symbol for
arbitrary polynomial in $\x^\mu$ but our wish is to work with
$L^2$ functions which are non-polynomial. For non-polynomial
functions one could in introduce the Weyl ordered symbol term by
term using the operator Taylor expansion,
\begin{equation}\label{op-Taylor}
  \hat{\mathcal{O}}(\x)=\mathcal{O}^{(0)}
  +\frac{1}{1!}\mathcal{O}_{\mu}^{(1)}\x^\mu+
  +\frac{1}{2!}\mathcal{O}_{\mu\nu}^{(2)}\x^\mu\x^\nu+\cdots.
\end{equation}

Although, application of the above procedure is possible in
principle for $L^2$ functions there is a more elegant way to find
the Weyl symbol which we are going to consider in the next
subsection.

\subsection{Noncommutative Calculus and the Star Product}
\label{sus:nc-calculus}

Let us introduce operators $p_\mu$ which are linear combinations
of $x^\mu$,
\begin{equation}\label{p}
  \p_\mu=-\theta^{-1}_{\mu\nu}\x^\mu.
\end{equation}
They have the following commutation relations,
\begin{equation}\label{p-comm}
  [\p_\mu,\p_\nu]=-\ii \theta^{-1}_{\mu\nu}.
\end{equation}

Consider adjoint operator,
\begin{equation}\label{P-adj}
  P_\mu=[\p_\mu,\cdot].
\end{equation}
This is operator acting on operators, i.e. it applies to elements
of $\A$: $P_\mu:\A\to\A$. It is not difficult to check that
$P_\mu$ are,
\begin{itemize}
 \item[i.] commutative,
\begin{equation}\label{P-comm}
  P_\mu P_\nu-P_\nu P_\mu=0;
\end{equation}
 \item[ii.] self-adjoint with respect to the scalar product defined
 on the algebra $\A$ of operators,
\begin{equation}\label{herm}
  \langle a, P_\mu b\rangle=\langle P_\mu a, b\rangle.
\end{equation}
The scalar product is defined as,
\begin{equation}\label{sc-prod}
  \langle a,b\rangle =\tr_\hh a^\dag b.
\end{equation}
 \item[iii.] They form a complete set of operators acting on $\A$,
in the sense that,
\begin{equation}\label{P-complete}
  \forall \mathbf{F}:P_\mu \mathbf{F}=0,(\mu=1,\dots,p) \Rightarrow
  \mathbf{F}=c\I,
\end{equation}
where $\I$ is the unity operator. This property follows from the
irreducibility of generators $\p_\mu$ (equivalently, $\x^\mu$).
\end{itemize}

Due to properties i.-iii. operators $P_\mu$ are diagonalizable and
having real eigenvalues. In particular, from property iii. it
follows that the dimension of common eigenspace of all $P_\mu$ is
one.

From the Leibnitz rule operators $P_\mu$ satisfy,
\begin{equation}\label{Leib}
  P_\mu(\mathbf{f}\cdot \mathbf{g})=P_\mu \mathbf{f}\cdot \mathbf{g}
  +\mathbf{f}\cdot P_\mu \mathbf{g},
\end{equation}
it follows that eigenfunctions $E_k$ of $P_\mu$ satisfy,
\begin{align}\label{eig}
  &P_\mu \mathbf{E}_k=k_\mu \mathbf{E}_k\\ \label{ee}
  &\mathbf{E}_k\cdot \mathbf{E}_{k'}\sim \mathbf{E}_{k+k'}.
\end{align}

From \eqref{ee} it is not difficult to guess that eigenfunctions
$E_k$ should have the form,
\begin{equation}\label{eigvect}
  \mathbf{E}_k=c_k \e^{\ii k_\mu \x^\mu}.
\end{equation}

Let us note that ``eigenvectors''  are elements of $\A$, i.e.
operators acting on the Hilbert space $\hh$.

As it can be seen, the spectrum of $P_\mu$'s is continuous and,
therefore, eigenvectors have infinite norm with respect to the
scalar product \eqref{sc-prod}. However, one can fix the quotients
$c_k$ from the requirement,
\begin{equation}\label{delta-norm}
  \tr \mathbf{E}_{k'}^\dag \cdot \mathbf{E}_k=\delta(k'-k).
\end{equation}

Let us compute the trace and find respective quotients. To do this
consider the basis where the set of operators $x^\mu$ splits in
pairs $p_i$, $q^i$ satisfying the standard commutation relations
\eqref{stand-Heis}.

As we know from courses of Quantum Mechanics the trace of the
operator
\begin{equation}\label{e.e}
  \e^{-\ii k'\x}\cdot \e^{\ii k\x}=\e^{\ii(k-k')\x}
  \e^{\frac{\ii}{2}k'\times k},
\end{equation}
can be computed in $q$-representation as,
\begin{equation}\label{qs1}
  \tr\e^{\ii(k-k')\x}
  \e^{\frac{\ii}{2}k'\times k}=\int\dd q \bra{q}
  \e^{-\ii(l'_i-l_i)q^i+(z^{\prime i}-z^i)p_i}\ket{q}=
  1/|c_k|^2\delta(k'-k),
\end{equation}
where $\ket{q}$ is the basis of eigenvectors of $\mathbf{q}^i$,
\begin{equation}\label{q-rep}
  \mathbf{q}^i\ket{q}=q^i\ket{q},\qquad
  \bracket{q'}{q}=\delta(q'-q),
\end{equation}
and $l_i$, $z^i$ ($l_i$, $z^i$) are components of $k_\mu$
($k'_\mu$) in the in the parameterizations: $x^\mu\to p_i,q^i$.
Explicit computation gives,
\begin{equation}\label{qs2}
  1/|c_k|^2=\frac{(2\pi)^{\frac{p}{2}}}{\sqrt{\det\theta}}.
\end{equation}

Now, we have the basis of eigenoperators $\mathbf{E}_k$ and can
write any operator $\mathbf{F}$ in terms of this basis,
\begin{equation}\label{F-expand}
  \mathbf{F}=\int\dd k\, \tilde{F}(k)\e^{\ii k\x},
\end{equation}
where the ``coordinate'' $\tilde{F}(k)$ is given by,
\begin{equation}\label{F-inv}
  \tilde{F}(k)=\frac{\sqrt{\det\theta}}{(2\pi)^{\frac{p}{2}}}
  \tr (\e^{-\ii k\x}\cdot \mathbf{F}).
\end{equation}

Function $\tilde{F}(k)$ can be interpreted as the Fourier
transform of an $L^2$ function $F(x)$,
\begin{equation}\label{Weyl-F}
  F(x)=\int \dd k \tilde{F}(k) \e^{\ii k_\mu x^\mu}=
  \sqrt{\det\theta}\int \frac{\dd k}{(2\pi)^{p/2}}\e^{\ii
  kx}\tr\e^{-\ii k\x}\mathbf{F}.
\end{equation}
Conversely, to any $L^2$ function $F(x)$ from one can put into
correspondence an $L^2$ operator $\mathbf{F}$ by inverse formula,
\begin{equation}\label{Weyl-inv}
  \mathbf{F}=\int \frac{\dd x}{(2\pi)^{p/2}} \int\frac{\dd k}{(2\pi)^{p/2}}F(x)
  \e^{\ii k(\x-x)}.
\end{equation}

Equations \eqref{Weyl-F} and \eqref{Weyl-inv} providing a
one-to-one correspondence between $L^2$ functions and operators
with finite trace,
\begin{equation}\label{tr}
  \tr \mathbf{F}^\dag\cdot \mathbf{F}<\infty,
\end{equation}
give in fact formula for the Weyl symbols. By introducing
distributions over this space of operators one can extend the
above map to operators with unbounded trace.
\begin{exercise}
 Check that \eqref{Weyl-F} and \eqref{Weyl-inv} lead in terms of
 distributions to \eqref{Weyl-poly}.
\end{exercise}

Let us note, that the map \eqref{Weyl-F} and \eqref{Weyl-inv} can
be rewritten in the following form,
\begin{equation}\label{f-delta}
  F(x)=(2\pi)^{p/2}\sqrt{\det\theta}\tr \dop (\x-x)
  \mathbf{F},\qquad \mathbf{F}=\int\dd^px\,\dop (\x-x) F(x),
\end{equation}
where we introduced the operator,
\begin{equation}\label{nc-delta}
  \dop (\x-x)=\int\frac{\dd^pk}{(2\pi)^p}\e^{\ii k\cdot(\x-x)}.
\end{equation}
This operator satisfy the following properties,
\begin{subequations}\label{dop-prop}
\begin{align}
  &\int \dd^px\, \dop(\x-x)=\I,\\
  &(2\pi)^{p/2}\sqrt{\det{\theta}}\tr \dop(\x-x)=1,\\
  &(2\pi)^{p/2}\sqrt{\det{\theta}}\tr \dop(\x-x)\dop(\x-y)=
  \delta(x-y),
\end{align}
where in the r.h.s. of last equation is ordinary delta function.
Also, operators $\dop (\x-x)$ form a complete set of operators if
regarded as a family depending on the parameter $x$,
\begin{equation}\label{dop-compl}
  [\dop (\x-x),\mathbf{F}]\equiv 0 \Rightarrow F\propto\I.
\end{equation}
The commutation relations of $\x^\mu$ also imply that $\dop(\x-x)$
should satisfy,
\begin{equation}\label{dop-comm}
  [\x^\mu,\dop(\x-x)]=\ii \theta^{\mu\nu}\pd_\nu\dop(\x-x).
\end{equation}
\end{subequations}

In fact one can define noncommutative plane starting from operator
$\mathbf{D}(x)$ satisfying \eqref{dop-prop}, with $\x^\mu$ defined
by,
\begin{equation}\label{x-mu}
  \x^\mu=\int\dd^px\, x^\mu \dop(\x-x).
\end{equation}
In this case \eqref{dop-comm} provides that $\x^\mu$ satisfy the
Heisenberg algebra \eqref{Heis1}, while the property
\eqref{dop-compl} provides that they form a complete set of
operators. Relaxing these properties allows one to introduce a
more general noncommutative spaces.

Let us the operator $\dop (x)$ in the simplest case of
two-dimensional noncommutative plane. The most convenient is to
find its matrix elements $D_{mn}(x)$ in the oscillator basis
\eqref{osc-bas},
\begin{equation}\label{dop-osc-bas}
  D_{mn}(x)=\bra{m}\dop^{(2)} (\a-z)\ket{n}=\tr \dop^{(2)}(\a-z)
  \mathbf{P}_{nm},
\end{equation}
where $\mathbf{P}_{nm}=\ket{n}\bra{m}$.

As one can see, up to a Hermitian transposition the matrix
elements of $\dop(\x-x)$ correspond to Weyl symbols of operators
like $\ket{m}\bra{n}$, or so called Wigner functions. The
computation of \eqref{dop-osc-bas} gives,\footnote{For the details
of computation see e.g. \cite{Harvey:2001yn}.}
\begin{equation}\label{D-matr-el}
  D^\theta_{mn}(z,\bar{z})=(-1)^n\left(\frac{2}{\sqrt{\theta}}\right)^{m-n+1}
  \sqrt{\frac{n!}{m!}} \e^{-z\bar{z}/\theta}
  \left(\frac{z^m}{\bar{z}^n}\right)L_n^{m-n}(2z\bar{z}/\theta),
\end{equation}
where $L_n^{m-n}(x)$ are Laguerre polynomials,
\begin{equation}\label{Lag}
  L_n^\alpha(x)=\frac{x^{-\alpha}\e^x}{n!}
  \left(\frac{\dd}{\dd x}\right)^n(\e^{-x}x^{\alpha+n}).
\end{equation}
It is worthwhile to note that in spite of singular looking
definition the symbol of the delta operator is a smooth function
rapidly vanishing at infinity. The smoothness comes from the fact
that the operator elements are written in an $L^2$ basis. In a
non-$L^2$ basis, e.g. in the basis of $x_1$ eigenfunctions
$D^{\theta}$ would have more singular form.

The above computations can be generalized to $p$-dimensions.
Written in the complex coordinates $z_i,\bar{z}_i$ corresponding
to oscillator operators \eqref{ai}, which diagonalize the
noncommutativity matrix this looks as follows,
\begin{equation}\label{D-matr-high}
  D_{\vec{m}\vec{n}}=D^{\theta_{(1)}}_{m_1n_1}(z_1,\bar{z}_1)
  D^{\theta_{(2)}}_{m_2n_2}(z_2,\bar{z}_2)
  \dots
  D^{\theta_{(p/2)}}_{m_{p/2}n_{p/2}}(z_{p/2},\bar{z}_{p/2}),
\end{equation}
where,
\begin{equation}\label{zz}
  [z_i,\bar{z}_i]_*=\theta_{(i)},\qquad i=1,\dots,p/2.
\end{equation}

Having the above map one can establish following relations between
operators and their Weyl symbols.
\begin{enumerate}
\item It is not difficult to derive that,
\begin{equation}\label{tr-int}
  (2\pi)^{p/2}\sqrt{\det\theta}\tr\mathbf{F}=\int \dd x\, F(x).
\end{equation}
\item The (noncommutative) product of operators is mapped into the
\emph{star} or \emph{Moyal} product of functions,
\begin{equation}\label{star}
  \mathbf{F}\cdot \mathbf{G}\to F*G(x),
\end{equation}
where $F*G(x)$ is defined as,
\begin{equation}\label{star-def}
  F*G(x)=\left.\e^{-\frac{\ii}{2}\theta^{\mu\nu}\pd_\mu\pd'_\nu}
  F(x)G(x')\right|_{x'=x}.
\end{equation}
In terms of operator $\dop(\x-x)$, this product can be written as
follows,
\begin{equation}\label{star-delta}
  F*G(x)=\int\dd^py\dd^pz\, K(x;y,z)F(y)G(z),
\end{equation}
where,
\begin{multline}\label{K}
  K(x;y,z)=\\
  (2\pi)^{p/2}\sqrt{\det{\theta}}\tr
  \dop(\x-x)\dop(\x-y)\dop(\x-z)=\\
  \e^{\frac{\ii}{2}\pd^y_\mu\theta^{\mu\nu}\pd^z_\nu}
  \delta(y-x)\delta(z-x),
\end{multline}
$\pd^y_\mu$ and $\pd^z_\mu$ are, respectively, $\pd/\pd y^\mu$ and
$\pd/\pd z^\mu$, and in the last line one has ordinary delta
functions.

From the other hand the ordinary product of functions was not
found to have any reasonable meaning in this context.

\item One property of the star product is that in the integrand
one can drop it once because of,
\begin{equation}\label{*-drop}
  \int \dd^px\, F*G(x)=\int \dd^px\,F(x)G(x),
\end{equation}
were in the r.h.s the ordinary product is assumed.

\item Interesting feature of this representation is that partial
derivatives of Weyl symbols correspond to commutators of
respective operators with $\ii \mathbf{p}_\mu$,
\begin{equation}\label{deriv}
  [\ii \mathbf{p}_\mu,\mathbf{F}]\to \ii(p_\mu*F-F*p_\mu)(x)
  =\frac{\pd F(x)}{\pd x^\mu},
\end{equation}
where $p_\mu$ is linear function of $x^\mu$:
$p_\mu=-\theta^{-1}_{\mu\nu}x^\nu$.
\end{enumerate}

This is an important feature of the star algebra of functions
distinguishing it from the ordinary product algebra. In the last
one can not represent the derivative as an \emph{internal
automorphism} while in the star algebra it is possible due to its
nonlocal character. This property is of great importance in the
field theory since, as it will appear later, it is the source of
duality relations in noncommutative gauge models which we turn to
in the next section.

\begin{exercise}
 Derive equations \eqref{tr-int}--\eqref{deriv}.
\end{exercise}
\newpage
\section{Noncommutative Field Theory}
As we mentioned in the Introduction the setup of noncommutative
geometry allows one to introduce and successfully develop the
notion of Noncommutative Field Theory. As we can anticipate from
what we learned in the previous section, one can introduce the
fields in the noncommutative space the same way as in ordinary
spaces except that the ordinary products should be replaced with
star products.

Here we will consider the classical aspects of Noncommutative
Field Theory. There are, however, many interesting things found in
the \emph{Noncommutative Perturbative QFT}, which we are not going
to discuss here but refer the Reader to the literature
\cite{Minwalla:1999px,VanRaamsdonk:2000rr}.

Another aspect is related to whether the \emph{time} is
commutative or not. When time can be chosen commutative one can in
principle define a canonically quantized theory. In the case of
noncommutative time one can not speak even on that. However, one
can work with path integrals in Euclidean time.

In classical analysis these subtleties are not important since
there one can easily pass from one case to another.

\subsection{Noncommutative Scalar Field}

The simplest field  theory model is one of the real scalar field.
It is given by the classical action,
\begin{equation}\label{sc-comm}
  S_{\mathrm{comm}}=\int d^px\,
  \left(\frac{1}{2}\pd_\mu\phi\pd_\mu\phi+V(\phi)\right),
\end{equation}
where $V(\phi)$ is a polynomial potential,
\begin{equation}\label{pot-com}
  V(\phi)=\frac{1}{2}m^2\phi^2+\frac{g_{(3)}}{3!}\phi^3+\dots.
\end{equation}
If $m^2>0$ than this defines a massive self-interacting scalar
model, while $m^2\equiv -\mu^2<0$ corresponds to Higgs models.

The generalization of above to the noncommutative case is
straightforward. The only difference which arise is the
substitution of ordinary products in \eqref{sc-comm} by star
products,
\begin{equation}\label{sc-nc}
  S_{\mathrm{nc}}=\int d^px\,
  \left(\frac{1}{2}\pd_\mu\phi*\pd_\mu\phi-V_*(\phi)\right),
\end{equation}
where $V_*(\phi)$ is the noncommutative interaction potential,
\begin{equation}\label{pot-nc}
  V_*(\phi)=\frac{1}{2}m^2\phi*\phi+
  \frac{g_{(3)}}{3!}\phi*\phi*\phi+\dots.
\end{equation}
Using the property \eqref{*-drop} one can drop the star products
from the quadratic free parts of the noncommutative action. Thus,
the noncommutativity arises only at the interaction level.

Let us try to rewrite the noncommutative action in the operator
form using the Weyl map \eqref{Weyl-F} and \eqref{Weyl-inv}. Under
this map one has to substitute the real field $\phi(x)$ by a
Hermitian operator $\hat{\phi}$
\begin{equation}\label{df->op}
  \pd_\mu\phi\to \ii[\mathbf{p}_\mu,\hat{\phi}].
\end{equation}
As a result one has the action in the operator form,
\begin{equation}\label{op-act}
  S_{\mathrm{nc}}=(2\pi)^{p/2}\sqrt{\det\theta}\tr\left(-\frac{1}{2}
  [\mathbf{p}_\mu,\hat{\phi}]^2+\mathbf{V}(\hat{\phi})\right),
\end{equation}
where the potential $\hat{V}(\hat{\phi})$ is inherited from the
star potential \eqref{pot-nc},
\begin{equation}\label{pot-op}
  \mathbf{V}(\hat{\phi})=\frac{1}{2}m^2\hat{\phi}\cdot \hat{\phi}+
  \frac{g_{(3)}}{3!}\hat{\phi}\cdot \hat{\phi}\cdot \hat{\phi}
  +\dots.
\end{equation}

As one can see the all dependence of the model on the data of
noncommutative space are now stored in the factor
$\sqrt{\det\theta}$ in front of the action and in operators
$\mathbf{p}_\mu$. By rescaling,
\begin{equation}\label{rescale}
  \hat{\phi}\to(2\pi)^{p/4}(\det{\theta})^{-\frac{1}{4}}\hat{\phi},
\end{equation}
the $\theta$-dependence reappears only in the interaction part of
the action.

As we see the operator form of the action is more invariant,
therefore it should be more fundamental. Later we will see that it
is indeed so.
\subsection{Gauge Invariance and Gauge Fields}

In ordinary field theories real singlet scalar field possesses no
special symmetries. The things are different, however, in the
noncommutative theory.

As we have mentioned in subsection \ref{unit-eq} the
noncommutative algebra \eqref{Heis1} allows a class of unitary
equivalent representations rather one single representation and it
would be dubitable why the model shall depend on the particular
representation. The change of representation is equivalent to
unitary transformation of all operators of the theory,
\begin{equation}\label{unit-tr}
  \mathbf{F}\to \mathbf{U}^{-1}\cdot\mathbf{F}\cdot\mathbf{U},
\end{equation}
for some unitary operator $U$,
\begin{equation}\label{U}
  \mathbf{U}^{\dag}\cdot\mathbf{U}=\I.
\end{equation}
In terms of Weyl symbols this means that the scalar field
$\phi(x)$ corresponding to the operator $\hat{\phi}$ undergoes the
transformation,
\begin{equation}\label{*-unit-tr}
  F(x)\to \bar{U}*F*U(x),
\end{equation}
where $U(x)$ is the respective star unitary function,
\begin{equation}\label{*U}
  \bar{U}*U=1,
\end{equation}
the bar denotes complex conjugation.

The equation \eqref{*-unit-tr} indicates that $\phi(x)$ transforms
in adjoint representation of noncommutative U(1) gauge group. If
we try to find global gauge transformations with $U=\e^{\ii
\alpha}=\const$ we find that they act trivially on $\phi(x)$. From
the other hand global gauge transformation correspond in the
operator language to constant phase transformation of the Hilbert
space vectors,
\begin{equation}\label{global-gt}
  \ket{\psi}\to\e^{\ii \alpha}\ket{\psi}.
\end{equation}
Therefore the scalar field action \eqref{sc-nc} and \eqref{op-act}
is obviously invariant with respect to these global
transformations, however it fails to be invariant with respect to
local transformations \eqref{*-unit-tr} due to noninvariance of
the kinetic term.

In ordinary theory one can extend the global gauge invariant model
to be local invariant by gauging the kinetic term. This is
obtained by the substitution of all partial derivatives of fields
with \emph{covariant derivatives} built of \emph{gauge fields}.
The same gauging procedure can be applied in noncommutative case
\cite{Sochichiu:2000rm}. It consists in the substitution of the
action \eqref{sc-nc} by the gauged one and addition to it of the
pure gauge field part $S_{\mathrm{g}}$,
\begin{equation}\label{gauging}
  S_{\mathrm{nc,gauged}}=\int d^px\,
  \left(\frac{1}{2}\nabla_\mu\phi*\nabla_\mu\phi-V_*(\phi)\right)
  +S_{\mathrm{g}}.
\end{equation}
The covariant derivatives,
\begin{equation}\label{cov-der}
  \nabla_\mu\phi=\pd_\mu\phi+\ii [A_\mu,\phi]_*(x)\equiv
  \pd_\mu\phi(x)+\ii(A_\mu*\phi-\phi*A_\mu)(x)
\end{equation}
imply that the gauge field transforms under the action local gauge
transformation as follows,
\begin{equation}\label{A-transf}
  A_\mu(x)\to U^{-1}*A_\mu*U(x)-\ii U^{-1}*\pd_\mu U(x).
\end{equation}
One can see, that due to the noncommutativity we got a nonabelian
group of local U(1) transformations for the gauge field as well.
Therefore the gauge part of the action should be constructed
respecting the gauge transformations of the gauge field. In this
case it takes the form,
\begin{equation}\label{gf-act}
  S_{\mathrm{g}}=-\frac{1}{4g^2}\int\dd^px\,F_{\mu\nu}*F^{\mu\nu},
\end{equation}
where,
\begin{equation}\label{f-mn}
  F_{\mu\nu}=\pd_\mu A_\nu-\pd_\nu A_\mu-\ii[A_\mu,A_\nu]_*.
\end{equation}

\subsection{Background Independence}

Let us now rewrite the gauged action in the operator form. In
order to do this let us observe that covariant derivatives of the
scalar fields can be represented as,
\begin{equation}\label{cov-op}
  \nabla_\mu\phi=\ii
  [(p_\mu+A_\mu),\phi]_*\to\ii[\mathbf{X}_\mu,\hat{\phi}],
\end{equation}
where $\mathbf{X}_\mu$ is operator corresponding to the function
$X_\mu(x)=p_\mu+A_\mu$. At the same time the gauge field strength
can be rendered as,
\begin{equation}\label{f-op}
  F_{\mu\nu}-\theta^{-1}_{\mu\nu}=[(p_\mu+A_\mu),(p_\nu+A_\nu)]_*
  \to[\mathbf{X}_\mu,\mathbf{X}_\nu].
\end{equation}

As we see covariant derivatives and
$F_{\mu\nu}-\theta^{-1}_{\mu\nu}$ can be represented by operators
independent of the generator basis $x^\mu$. In fact, as
$\theta_{\mu\nu}$ is constant the gauge field action is
indistinguishable in what concerns equations of motion from one
where all $F_{\mu\nu}$ are substituted by
$F_{\mu\nu}-\theta^{-1}_{\mu\nu}$,
\begin{equation}\label{sg-mod}
  S'_{\mathrm{g}}=-\frac{1}{4g^2}\int\dd^px\,
  (F_{\mu\nu}-\theta^{-1})^{2}=-(2\pi)^{p/2}
  \sqrt{\det{\theta}}\frac{1}{4g^2}
  \tr[\mathbf{X}_\mu,\mathbf{X}_\nu]^2.
\end{equation}

Combining all together we can write down the action of
noncommutative gauge model of the scalar field in the operator
form,
\begin{equation}\label{sc-op}
  S=-\tr\left(\frac{1}{4{g'}^2} [\mathbf{X}_\mu,\mathbf{X}_\nu]^2+
  \frac{1}{2}[\mathbf{X}_\mu,\hat{\phi}]^2
  +\mathbf{\tilde{V}}(\hat{\phi})\right),
\end{equation}
were we introduced modified couplings,
\begin{align}
  &\mathbf{\tilde{V}}(\hat{\phi})=
  \frac{1}{2}m^2\hat{\phi}\cdot \hat{\phi}+
  \frac{g'_{(3)}}{3!}\hat{\phi}\cdot \hat{\phi}\cdot \hat{\phi}
  +\dots,\\
  &g'=\frac{g}{(2\pi)^{\frac{p}{4}}(\det{\theta})^{\frac{1}{4}}},\\
  &g'_{(n)}=(2\pi)^{\frac{p(2-n)}{4}}
  (\det{\theta})^{\frac{2-n}{4}}g_{(n)}.
\end{align}

There is no difficulty to generalize above to the case of a
multiplet of scalar fields. Consider now the particular case when
there is a scalar multiplet $\phi_a(x)$ $a=1,\dots,n$, with the
potential
\begin{equation}\label{Higgs-pot}
  V_{\mathrm{H}}(\phi)=-\frac{g^2}{4}[\phi_a,\phi_b]^2_*.
\end{equation}
This potential has a valley of nontrivial vacua and it plays an
important role in the dynamics of branes. (The fields $\phi_a(x)$
are believed to describe the transversal degrees of freedom of
branes, while gauge fields are responsible for the longitudinal
ones.) This gauge model has a fairly simple operator form,
\begin{equation}\label{Higgs-op}
  S_{\mathrm{H}}=-\frac{1}{4g^2}\tr[\mathbf{X}_M,\mathbf{X}_N]^2,
\end{equation}
where,
\begin{equation}\label{X}
  \mathbf{X}_M=
  \begin{cases}
    X_\mu, & M=\mu,\\
    \phi_a, & M=a;
  \end{cases}\qquad M=1,\dots,p+n.
\end{equation}

\begin{exercise}
  Prove \eqref{Higgs-op} and \eqref{X}.
\end{exercise}

As one can see the action in the form \eqref{sc-op} do not depend
explicitly on generators $\p_\mu$ or $\x^\mu$. The only input required
is the Hilbert space of representation generated by them. Thus, if a
different algebra can be represented on the same Hilbert space this
equation could equally apply to it. Algebras which can have exact
representations on the same Hilbert space are called \emph{Morita
equivalent}. In fact all Heisenberg algebras \eqref{Heis1} are Morita
equivalent because all infinite-dimensional separable Hilbert spaces
are isomorphic.  The isomorphism follows from the existence of
countable basis in each separable Hilbert space.

In section \ref{sec:nc-dual} we will explore this fact to show duality
relations arising between different gauge models. Before turning to it
let us consider the topic of Wilson lines in noncommutative gauge
models.

\textbf{Note:} Background invariance defined by us here is
different from one discussed e.g. in \cite{Seiberg:2000zk}. The
difference consists in the interpretation of $X_\mu$ as
derivatives rather than noncommutative coordinates. Although, for
$[X,X]=$constant background this reduces simply to redefinition
$X_\mu\to\theta^{\mu\nu}X_\nu$ or, equivalently respective
redefinition of the metric $g_{\mu\nu}$ which contracts $X_\mu$,
for $[X,X]\neq$constant these two possible choices are rather dual
then equivalent. The noncommutative space can be naturally defined
in terms of derivatives or vector fields on it.
\subsection{Wilson lines}
Wilson lines and Wilson loops in context of noncommutative gauge
theory or matrix models where consi9dered in Refs.
\cite{Ishibashi:1999hs,Alekseev:2000td,Das:2000md,Abou-Zeid:2000qi,%
Dhar:2000nj,Das:2000ur,Dhar:2001ff,Lee:2001sz,Dhar:2001ba,%
Bassetto:2001is,Kiem:2001fn}.

In commutative gauge theories Wilson loops or closed Wilson lines
play an important role of local gauge invariant observables. The
generic Wilson line in a nonabelian ordinary (i.e. on a
commutative space) Yang--Mills theory is defined as path ordered
exponent,
\begin{equation}\label{com-wl}
  U[C](x,y)=P\exp \int_{C}  A_\mu (x)\dd x^\mu,
\end{equation}
where $C$ is a line connecting points $x$ and $y$, e.g.
parameterized as $x^\mu(t)$: $x^\mu(0)=x^\mu$, $x^\mu(1)=y^\mu$.
The $P$-symbol in front of the exponent means that the exponential
in r.h.s. of eq. \eqref{com-wl} is computed by multiplying factors
with bigger parameter $t$ to the right of those with smaller $t$.

Under an ordinary nonabelian gauge transformation $A\to
g^{-1}Ag+g^{-1}\dd g$ the Wilson line transforms in the bi-local
manner,
\begin{equation}\label{wl-gt}
  U[C](x,y)\to g^{-1}(x)U[C](x,y)g(y).
\end{equation}

Thus, if one takes $x=y$ i.e. $C$ to represent a closed loop then one
can make out of $U[C]$ a gauge invariant quantity,
\begin{equation}\label{wl-inv}
  W[C](x)=\tr U[C](x,x),
\end{equation}
where the trace is performed over the gauge group.

The Wilson loop observables \eqref{wl-inv} play an important role
in analyzing the phase structure of the ordinary four dimensional
nonabelian Yang--Mills model as the order parameter. From the
other hand locally time-like (in Minkowski space) Wilson loop is
the interaction potential between a charged particle-antiparticle
pair e.g. interacting quark and antiqark. The behavior of the
expectation value of $W[C]$ describes the character of interaction
of opposite charges in the theory.

Thus in particular if the Wilson loop expectation value for large
loops has the behavior $\langle W[C]\rangle \sim \exp(-A(C))$,
where $A(C)$ is the minimal area spanned by the loop $C$, then it
is not difficult to see that the $q\bar{q}$ interaction potential
increases linearly which means confinement. This is called
\emph{area-law behaviour} of Wilson loop expectation value. It is
realized in the strong coupling regime of commutative Yang--Mills
model.

Another interesting regime is described by the \emph{perimeter
law} behavior i.e. $\langle W[C]\rangle \sim \exp(-L(C))$, where
$L(C)$ is the perimeter of the loop. That is realized in the weak
coupled limit. The perimeter low indicates that the interaction is
Coulomb-like.

For Abelian theories one of course does not need to write path
ordering and trace to define the Wilson loop invariant. The
particular meaning of the Wilson loop invariant here is the
electro-magnetic flux through $C$. Therefore, \eqref{wl-inv} can
be interpreted as definition of the analog of the field strength
flux in the nonabelian theory.

Let us turn to the noncommutative Yang--Mills model. Formally, one
can apply the rule given by commutative formula \eqref{wl-gt} for
the Weyl symbols to get,
\begin{equation}\label{wl-nc}
  U_*[C](x,y)=P\exp_*\int \dd x^\mu A_\mu (x),
\end{equation}
where $\exp_*$ means that the exponential is computed using the
star product \eqref{star-def},
\begin{equation}\label{*exp}
  \exp_* f(x)=1+\frac{1}{1!}f(x)+\frac{1}{2!}f*f(x)+\dots
\end{equation}

Due to noncommutativity of the star product we had to impose path
ordering even for U(1) gauge group. The quantity \eqref{wl-nc}
transform ``bi-locally'' under star gauge transformations analogously
to \eqref{wl-gt}. Therefore to make a gauge invariant object out of
closed loops we have to trace them out. Recall, however that the
tracing in noncommutative theory is equivalent to
integration. Therefore, invariant Wilson loop invariants in
noncommutative gauge theories look like,
\begin{equation}
 W[C]=\tr U[C]\propto\int\dd x \, U[C](x,x).
\end{equation}

As we see not much of $x$-dependence remains in the gauge invariant
objects constructed out of the Wilson loops. Indeed, we could not
expect to have a set of gauge invariant local functions since any
local function must obey adjoint transformation rule under the gauge
group action!

Another problem is that the above definition of the Wilson line
runs in terms of functions rather then in the terms of operators
which we are inclined to attribute more fundamental role.

\begin{figure}[t]
 \centering{\input{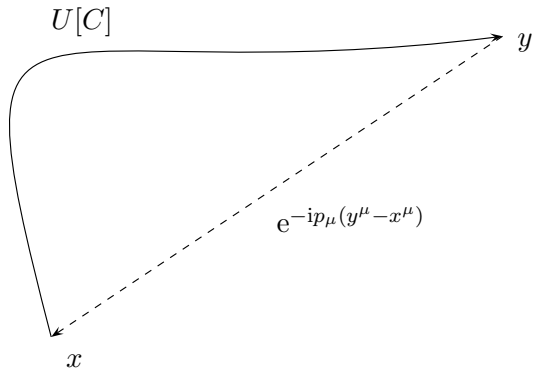}}
 \caption{An ``open'' Wilson line in noncommutative space. The
dashed line corresponds to insertion of the translation operator
$\e^{-\ii p_\mu \Delta x^\mu}$ from the end to the beginning of
the line.}\label{fig:Wilson}
\end{figure}
From the other hand, in noncommutative case one can multiply an
open Wilson line \eqref{wl-nc} by the function $\e^{-\ii
p\cdot(\Delta x)}$, where $\Delta x^\mu=y^\mu-x^\mu$ as shown in
the Fig.\ref{fig:Wilson}. Since $\e^{-\ii p\cdot(\Delta x)}$ is
the translation operator from $y$ to $x$ the modified Wilson line
will transform under the gauge group action as a ``local''
noncommutative function of $x$. Therefore, integrating it out will
produce a gauge invariant quantity,
\begin{equation}\label{wl-mod}
  W'[C]=\int \dd^p x\,\left(P\e_*^{\ii\int_C A_\mu(x)\dd x^\mu}
  \e^{-\ii p_\mu\Delta x^\mu}\right).
\end{equation}

Definition of Wilson lines \eqref{wl-nc} strongly depends on the
background and therefore there is no indication that Wilson lines
should be background independent. This is indeed so for the open
Wilson line \eqref{wl-nc}. However, it surprisingly appears that
Wilson line with the shifted end \eqref{wl-mod} is background
invariant. (When it is multiplied with the $\theta$-dependent
factor, $(2\pi)^{p/2}\sqrt{\det{\theta}}$.)

To show this, consider first a straight Wilson line $C_{\Delta
x}$. In this case, the Wilson line with shifted end acquire the
following simple operator form,
\begin{equation}\label{wl-str}
  W'[C_{\Delta x}]=(2\pi)^{\frac{p}{2}\sqrt{\det{\theta}}}
  \tr \e^{\ii\Delta x^\mu X_\mu}.
\end{equation}

The equivalence of \eqref{wl-str} and \eqref{wl-mod} for straight
lines can be proved in the following way. Let us divide the line
$C_{\Delta x}$ in a large number $N$ of equal pieces. Then, one
can split the exponential factor in \eqref{wl-str} in a product of
$N$ factors,
\begin{multline}\label{prod}
  W[C_{\Delta x}]\sim \tr\e^{\ii \Delta^\mu X_\mu}=\\
  \tr \prod_{n}\e^{\ii \Delta^\mu_{(n)} X_\mu}=
  \tr\prod_{n}\e^{\ii \Delta^\mu (\p_\mu+\mathbf{A}_\mu)}=\\
  \tr (1+\ii\Delta^\mu \mathbf{A}_\mu^{(1)})\dots
  (1+\ii\Delta^\mu \mathbf{A}_\mu^{(n)})\dots
  (1+\ii\Delta^\mu \mathbf{A}_\mu^{(N)})\\
  \times(1+\ii\Delta^\mu \p_\mu)^N+O(N^{-1}),
\end{multline}
where each $\Delta^\mu_{(n)}=\Delta x^\mu/N$, and
\begin{equation}\label{a-trans}
  \mathbf{A}^{(n)}_\mu=(1+\ii\Delta^\mu \p_\mu)^n \mathbf{A}_\mu
  (1-\ii\Delta^\mu \p_\mu)^n.
\end{equation}

Rewriting \eqref{prod} and \eqref{a-trans} in the Weyl form and
taking the limit $N\to\infty$, one gets precisely equation
\eqref{wl-mod} for the straight line. In particular, operator
$\mathbf{A}^{(n)}_\mu$ defined in Eq. \eqref{a-trans} maps to
$A_\mu (x(t))$, where $t=\lim_{N\to\infty}n/N$.

Equation \eqref{wl-str} is not yet background invariant, but can
be easily made so. The background invariance of $W'[C_{\Delta x}]$
is affected by the $\theta$-dependent factor in the r.h.s. of Eq.
\eqref{wl-str} and by the the explicit dependence of the $X_\mu$
field only. Therefore, the generalization to background invariant
form looks as,
\begin{equation}\label{wl-str-bg}
  W[C_{u}]=\tr \e^{\ii u^i X_i},\qquad i=1,\dots,D.
\end{equation}

Now, let us generalize the background invariant formula
\eqref{wl-str-bg} to arbitrary smooth Wilson lines, not
necessarily straight ones. Any smooth Wilson line can be imagined
as consisting of infinitesimal straight lines. The observable
corresponding to entire line corresponds to the product of
straight factors,
\begin{equation}\label{wl-frak}
  W[C]=
  \tr \e^{\ii u^i_1X_i}\e^{\ii u^i_2X_i}\dots
  \e^{\ii u^i_NX_i},\qquad C=C^{\mathrm{stright}}_1C^{\mathrm{stright}}_2\dots
  C^{\mathrm{stright}}_N,
\end{equation}
whose smooth limit is the path-ordered integral,
\begin{equation}\label{wl-gen}
  W[C]=\tr P_u\e^{\ii\int_C\dd u^i X_i},
\end{equation}
where ``$P_u$'' denotes ordering with respect to parameter $u$,
while $C$ denotes an arbitrary line in $\mathbb{R}^D$.

In a particular background $X_\mu=p_\mu$, $X_a=\const$ the line
$C$ maps to a line $C'$ in the $p$-dimensional space. Applying a
method similar to one given by eq. \eqref{prod}, one can obtain
the Weyl form of the generalized Wilson line \eqref{wl-gen} in
this background,
\begin{equation}\label{wl-mod-gen}
  W[C']=\frac{1}{(2\pi)^{\frac{p}{2}}\sqrt{\det{\theta}}}
  \int \dd^px\, \e_*^{\ii (\int_{C'} \dd x^\mu\,
  A_\mu+\int_{C'}\dd t\, \dot{u}^a(t)\phi_a)}.
\end{equation}
The line $C'$ in above equation is given by the projection of $C$
to the space spanned by $\p_\mu$, i.e. if $u^i(t)$, $0\leq t\leq
1$, $i=1,\dots,D$ defines line $C$, then $C'$ is defined by
equation $x^\mu(t)=u^\mu(t)$, $\mu=1,\dots,p$. As we see the line
$C$ in the auxiliary space $\mathbb{R}^{D}$ corresponds to a
$p$-dimensional line with a $(D-p)$-dimensional vector fibre over
it.

\newpage
\section{Noncommutative Gauge Dualities}\label{sec:nc-dual}
This section follows the original papers
\cite{Sochichiu:2000bg,Sochichiu:2000kz}.

As we have discovered above, the noncommutative gauge model with
scalar fields can be reformulated in the operator form where the
dependence of the noncommutative space enters only through the
representation of the noncommutative space algebra. Spanning
algebras corresponding to different noncommutative spaces but
having isomorphic representation will give us different equivalent
noncommutative models.

As we discussed in the section \ref{section:nc-space}, the
noncommutative settings are stored in three factors: the operator
algebra, its representation, and the (set of) derivative
operators. While representation and operator algebras are
generally fixed, the choice of derivative operators is at our
disposal. In fact we can switch from one noncommutative $\R$ to
another or from one gauge group to a different one by choosing
derivative with appropriate symmetries. This is possible because
particular gauge field backgrounds in some spaces can be
interpreted as a pure partial derivatives in different
noncommutative spaces. This equivalence relation is described by
the Seiberg--Witten map. (See Appendix \ref{app:swm}.)

Here we consider the cases of maps relating models in different
dimensions, and ones relating models with different gauge groups.
\subsection{Equations of Motion and Constant Curvature Solutions}
In this subsection we are interested mainly in the scalar
multiplet model with the potential \eqref{Higgs-pot}.

Gauge field equations of motion corresponding to the action with
potential \eqref{Higgs-pot} take the standard form,
\begin{subequations}\label{EOM-Wfj}
\begin{align}\label{EOM-W}
  &\nabla_\mu F_{\mu\nu}=-j_\nu,\\
  &\nabla^2\phi_a=0,
\end{align}
where
\begin{equation}\label{j}
  j_\mu=-\ii[\phi_a,\nabla_\mu\phi_a]_*,\text{ and
  }\nabla^2=\nabla_\mu\nabla^\mu.
\end{equation}
\end{subequations}
are, respectively, the noncommutative current generated by the
scalar fields and the noncommutative Laplace operator
respectively. In the operator form equations of motion have the
form,
\begin{equation}\label{EOM-inv}
  [\mathbf{X}_M,[\mathbf{X}_M,\mathbf{X}_N]]=0,\qquad M,N=(\mu,a),
\end{equation}
or, equivalently,
\begin{subequations}\label{EOM-op}
\begin{align}
  [\mathbf{X}_\mu,[\mathbf{X}_\mu,\mathbf{X}_\nu]]=-\mathbf{j}_\nu,
  &\qquad \mathbf{j}_\nu=[\phi_a,[\mathbf{X}_\mu,\phi_a]],\\
  [X_\mu,[X_\mu,\phi_a]]&=0.
\end{align}
\end{subequations}

Of course all three forms of equations of motion \eqref{EOM-Wfj},
\eqref{EOM-inv} and \eqref{EOM-op} are equivalent, the second
\eqref{EOM-inv} being also the most compact.

An obvious solution to the equations of motion is the constant
gauge field strength one,
\begin{equation}\label{cons-F}
  F^{(0)}_{\mu\nu}(x)=\const_{\mu\nu},\qquad\phi^{(0)}=\const.
\end{equation}
The operator form \eqref{EOM-inv} of the equations of motion
suggest a more general solution,
\begin{equation}\label{const-[XX]}
  [\mathbf{X}^{(0)}_M,\mathbf{X}^{(0)}_N]=-\ii\Theta_{MN}\I,
\end{equation}
which in terms of fields $F_{\mu\nu}$ and $\phi$ reads,
\begin{subequations}\label{sol-Fphi}
\begin{align}
 & F^{(0)}_{\mu\nu}=-\Theta_{\mu\nu},\qquad
 [\phi^{(0)}_a,\phi^{(0)}_b]=-\Theta_{ab},\\
 & \qquad\nabla_{\mu}\phi^{(0)}_a=-\Theta_{\mu  a},
\end{align}
\end{subequations}
where $\Theta_{MN}$ is split in blocks according to \eqref{X},
\begin{equation}\label{Theta-blocks}
  \Theta_{MN}=
  \begin{pmatrix}
  \Theta_{\mu\nu} & \Theta_{\mu b}\\
  \Theta_{a\nu}&\Theta_{ab}
  \end{pmatrix}.
\end{equation}

\subsection{Dimension Changing Solutions}
Let us see what is the effect of the solution
\eqref{const-[XX]}. First assume that this solution is given by a
\emph{complete} (in the sense of Quantum Mechanics) set of operators
$\mathbf{X}_M$ satisfying,
\begin{equation}\label{sol-compl}
  \forall \mathbf{F}:[\mathbf{X}_M,\mathbf{F}]=0, \qquad
  M=1,\dots,p+n \Rightarrow \mathbf{F}\propto\I.
\end{equation}

If $p'$ is the rank of $\Theta_{MN}$, then it has $D-p'$ zero
modes. Let us divide the indices  $M,N,\dots$ in primed early
roman indices $a', b'$ etc., which run in the zero space of
$\Theta_{MN}$ and primed Greek $\mu',\nu'$ etc., running in the
orthogonal completion. In other words, in primed indices introduce
the basis where $\Theta_{MN}$ have the following block structure:
\begin{equation}\label{Theta-blocks-prime}
  \Theta_{MN}=
  \begin{pmatrix}
  \theta'_{\mu'\nu'} & 0\\
  0&0
  \end{pmatrix}.
\end{equation}

Applying now the machinery developed by us in the subsection
\ref{sus:nc-calculus} we end up with a gauge model having a
different field content: $A_{\mu'}$, $\mu'=1,\dots,p'$ and
$\phi_{a'}$, $a'=1,\dots,D-p$. Such an operation can be applied to
any field participating in the gauge interaction in the adjoint
representation to give rise to a field in the new background. (As
we established, any real field, in fact, transforms in the adjoint
representation of the gauge group.) As a result any field $F(x)$
corresponding to a background invariant operator $\mathbf{F}$ in
the ``old'' theory in the ``new'' theory will be represented by a
field $F'(x')$, according to the formula,
\begin{equation}\label{FF'-transform}
  F'(x')=\sqrt{\frac{\det\theta'}{\det\theta}}
  \int\frac{\dd^{p'} k'}{(2\pi)^{p'/2}}\e^{\ii k'x'}
  \int\dd^{p} x\,\e_*^{-\ii {\theta'}^{\mu'\nu'}k'_{\mu'}
  p'_{\nu'}(x)}F(x),
\end{equation}
where ${\theta'}^{\mu'\nu'}$ is inverse matrix to
${\theta'}_{\mu\nu}$ and $p'_{\mu'}$ is the projection of the
solution $p_M(x)\equiv (p_\mu+A^{(0)}_\mu(x),\phi^{(0)}_a(x))$ to
the nonzero subspace of $\theta_{MN}$ (see eq.
\eqref{Theta-blocks-prime}).

\begin{exercise}
 Derive \eqref{FF'-transform} using the analysis of subsection
 \ref{sus:nc-calculus}.
\end{exercise}
\begin{exercise}
 Find transformation rules for fields $\phi_a$ and for gauge
 fields in the case of infinitesimal change of the solution,
 $\delta p_\mu=\epsilon_{\mu\nu}p_\nu$.
\end{exercise}

Thus, we established an interesting feature of the noncommutative
models involved in gauge interaction: \emph{the dimensionality of this
models can be changed switching out between fields $A_\mu$ and
$\phi_a$.}

The question one may still ask is whether such gauge field
configurations exist and how to construct them.

The existence of different gauge configurations satisfying
irreducibility condition \eqref{sol-compl} follows from the
isomorphism of separable infinite dimensional Hilbert spaces
realizing irreducible representation of Heisenberg algebras
generated by $X_M$. The isomorphism of infinite dimensional
Hilbert spaces is a result of separability. Let us remind that by
definition the separable infinite dimensional Hilbert space is one
having a countable basis. The isomorphism between such spaces is
realized by the identification of elements with same numbers in
countable bases of such spaces. It is clear that this isomorphism
is defined up to unitary transformations due to ambiguity in
choosing the Hilbert space basis.

Consider now, as an example, the two-dimensional
Yang--Mills--Higgs model with at least two scalar fields and let
us construct a solution in vicinity of which the theory is
effectively described by the four-dimensional model with two
scalar fields less. To construct such a field configuration in the
operator form, let us fix the isomorphic map $\sigma$ connecting
Hilbert spaces of two and four dimensional Heisenberg algebras.

The map $\sigma$ can be constructed as follows. Consider the
oscillator basis \eqref{p/2-osc} for $p=4$. and let us enumerate
the oscillator states. Denote the unique number assigned to each
state $\ket{\vec{n}}$ as $n(\vec{n})\equiv n(n_1,n_2)$. We have
then,
\begin{equation}\label{map}
  \sigma:\ket{\vec{n}}\mapsto \ket{n=n(\vec{n})},
\end{equation}
where $\ket{n}$ is the element of oscillator basis of
two-dimensional Heisenberg algebra. Since the number assigned to
the lattice vector is unique this map is isomorphic.

Any operator $\mathbf{F}_{(2)}$ acting on the Hilbert space of one
oscillator can be uniquely mapped to an operator
$\mathbf{F}_{(4)}$ acting on two oscillator Hilbert space, by the
rule
\begin{equation}\label{op-map}
  \mathbf{F}_{(2)}\mapsto \mathbf{F}_{(4)}=\sigma^{-1}\cdot \mathbf{F}_{(2)}\cdot
  \sigma.
\end{equation}
Let us remind that such operators $\mathbf{F}_{(4,2)}$ correspond
via Weyl map to, respectively, four-dimensional and
two-dimensional noncommutative functions. Thus, this is also an
isomorphism between spaces of these functions.

Since the map is realized in the operator form it preserves the
star algebra in the sense that it is a homomorphism of star
algebras, i.e.,
\begin{equation}\label{hom*}
  F_{(2)}*_{(2)}G_{(2)}\mapsto
  F_{(4)}*_{(4)}G_{(4)},
\end{equation}
where $*_{(2,4)}$ are respective star products on the two- and
four-dimensional noncommutative functions $F_{(p)}(x_{(p)})$,
where the subscript ${}_{(p)}$, denote the dimension (in our case
it is $p=2$ or $p=4$). Let us note, that the data about the
noncommutativity parameter $\theta^{\mu\nu}$ are contained in the
definition of oscillator operators $\a_i$ and $\bar{\a}_i$.

The above construction was performed in the operator formalism.
When trying to pass to the Weyl symbols one faces the problem that
expressions defining the Weyl symbols are not valid for such
operators as $\p_\mu$ or $\p'_{\mu'}$. Indeed, it is not difficult
to see, that at least half of operators satisfying the Heisenberg
algebra should have divergent trace, and therefore divergent
integral in the Fourier transform. However, as we know, in the
theory of commutative functions one can extend the definition of
the Fourier transform to non square integrable functions if to
work in terms of distributions rather than in that of ordinary
functions.

As we discussed in the section \ref{section:nc-space}, in the
noncommutative case one can also introduce the notion of a
``noncommutative'' linear distribution. Let us rewrite the map
\eqref{FF'-transform} in terms of distribution-valued operators,
\begin{equation}\label{dops}
  \mathbf{D}(x)=\dop(\x-x),\text{ and }\mathbf{D}'(x')=\dop(\x'-x').
\end{equation}

Then, the map \eqref{FF'-transform} can be rewritten as,
\begin{equation}\label{FF'-delta}
  F'(x')=(2\pi)^{p'/2}\sqrt{\det{\theta'}}\int \dd^px\, D(x',x)F(x),
\end{equation}
where,
\begin{equation}\label{D(xx)}
  D(x',x)=\tr\mathbf{D}'(x')\mathbf{D}(x)=
  \sum_{\vec{m},\vec{n}} D'_{\vec{m}'(\vec{m})\vec{n}'(\vec{n})}(x')
  D_{\vec{n},\vec{m}}(x).
\end{equation}
This can be represented as follows,
\begin{equation}\label{v}
  D(x',x)=
  \sum_{\ov{\vec{m},\vec{n}}{\vec{m}',\vec{n}'}}
  D'_{\vec{m}'\vec{n}'}(x')V_{\vec{n}'\vec{n}}
  D_{\vec{n},\vec{m}}(x)V^\dag_{\vec{m}\vec{m}'},
\end{equation}
where the unitary operator $\mathbf{V}$ defined by,
\begin{equation}
  \mathbf{V}=\sum_{\vec{n}',\vec{n}}V_{\vec{n}'\vec{n}}
  \ket{\vec{n}'}\bra{\vec{n}},
\end{equation}
can be viewed as the map from representation of $p$-dimensional
Heisenberg algebra to the space of representation of the
$p'$-dimensional one. The operator $\mathbf{V}$ has the meaning of
element of the equivalence bimodule $\hh_{(p')}\otimes\hh_{(p)}$
(see Appendix \ref{app:kt}) realizing Morita equivalence of two
algebras. Heisenberg algebras are trivial from the K-theory point
of view since any module is an infinite-dimensional separable
Hilbert space and it is isomorphic to the space of operators with
bounded trace of squares (see \eqref{tr}) which is an
infinite-dimensional separable Hilbert space too.

\subsection{Gauge Group Changing Solutions}

Let us return to the solution \eqref{const-[XX]}, but relax the
condition \eqref{sol-compl}. Let, now operators
$\mathbf{X}_M=\{\p_\mu,0\}$ fail to form a complete set of
operators, i.e. there are such operators $\F_a$, which commute
with all $\p_\mu$,
\begin{equation}\label{deg}
  [\p_\mu,\F_a]=0,\qquad \mu=1,\dots,p,
\end{equation}
but which are not scalar, $\F_a\neq c_a\I$. Let a set of $\F_a$,
$a=1,\dots,n$ be chosen in such a way that the total set
consisting of both $\p_\mu$ and $F_a$ is complete, i.e.
\begin{equation}\label{F-complete}
   \forall \mathbf{F}:[p_\mu,\mathbf{F}]=
  [\F_a,\mathbf{F}]=0 \Rightarrow
  \mathbf{F}=c\I.
\end{equation}
(Generically, one can always supplement the set of operators
$\p_\mu$ by some other operators $\F_a$, such that the total set
to be complete.)

As it can be seen, from \eqref{deg} and \eqref{F-complete} it
follows that operators $\F_a$ should form an algebra,
\begin{equation}\label{F-alg}
  [\F_a,\F_b]=\ii G_{ab}(\F),
\end{equation}
where $C_{ab}$ is an operator function of $\F_a$. In particular,
it can be linear in $\F_a$, like
\begin{equation}\label{Lie}
  C_{ab}(\F)=C_{ab}^c\F_c,
\end{equation}
where $C_{ab}^c$, are structure constants of a (semi)simple Lie
algebra $\mathcal{G}=Lie(G)$. In the last case operators $\F_a$
should form an irreducible representation of the respective Lie
algebra.

Let us show that the model around such a background looks like
noncommutative gauge model with the gauge group $\U(1)\times G$.
To do this one should perform the analysis similar to one of the
subsection \ref{sus:nc-calculus}.

The difference of the present case from the standard one analyzed
in subsection \ref{sus:nc-calculus} is that now the set of
operators $\p_\mu$ is not complete and so is the set of adjoint
operators $P_\mu$. Therefore, the eigenvalues of $P_\mu$ become
degenerate.

Since operators $\F_a$ commute with $\p_\mu$ they also should
commute with $P_\mu$. This means that the eigenspace of $P_\mu$
corresponding to a particular momentum $k$ is invariant under
action of $\F_a$. Therefore, this eigenspace realizes a
representation of the algebra \eqref{F-alg}. Moreover, in virtue
of the Schur's lemma this representation is irreducible, it is the
adjoint representation of the algebra $u(1)\oplus \mathcal{G}$.
The factor $u(1)$ comes from the fact that the unity operator is
always present in the algebra of operators. It corresponds to
operators which are ``singlet'' in $\mathcal{G}$, i.e. commute
with all $\F_a$. They, are functions of $\p_\mu$ and, therefore,
are not trivial.

As a result, we have that the Hilbert space $\hh$ can be split
into a tensor product as follows,
\begin{equation}\label{tens-prod}
  \hh\simeq\hh'\otimes V_{\mathcal{G}},
\end{equation}
where $\p_\mu$ are irreducible on $\hh'$, while $\mathcal{G}$ on
$V_{\mathcal{G}}$.

Obviously, $\hh'\simeq \hh$, which means that the infinite
dimensional separable Hilbert space is isomorphic to itself times
a Hermitian space. Let us construct this isomorphic map for a
particular case of $\mathcal{G}=\su(2)$. For this, fix two bases
in, respectively, $\hh$ and $\hh\otimes V_{(2)}$ where $V_{(2)}$
is two-dimensional Hermitian space,
\begin{align}
  &\ket{n}\in \hh, &n=0,1,2,\dots,\\
  &\ket{n}\otimes e_\alpha\in\hh\otimes V_{(2)},
  &n=0,1,2,\dots,\qquad \alpha=0,1,
\end{align}
$\{e_\alpha,\alpha=0,1\}$ is the basis of $V_{(2)}$. The map is
obtained by the identification of the subspace of $\hh$
corresponding to even values of $n$ to the subspace $\hh\otimes
e_0$ of $\hh\otimes V_{(2)}$, and respectively, the subspace with
odd values of $n$ is mapped to $\hh\otimes e_1$. On the basis
elements it looks as follows,
\begin{equation}\label{map-2}
  \sigma:\ket{n}\mapsto \ket{[n/2]}\otimes e_\alpha,
\end{equation}
where $\alpha=n \mod 2$, and $[n/2]$ is the integer part of $n$.

As above, with the map $\sigma$ at hand we can pull back any
noncommutative function from the ``nonabelian'' $\U(1)\times G$
Yang--Mills--Higgs theory to the $\U(1)$ Yang--Mills--Higgs model,
\begin{equation}\label{pull-un}
  \mathbf{F}_{U(1)}=\sigma^{-1}\cdot\mathbf{F}_{U(n)}\cdot\sigma,
\end{equation}
and vice versa using $\sigma^{-1}$.

An analysis in terms of distributions analogous to one carried in
the previous subsection can be also performed here.
\begin{exercise}
  Perform the analysis of subsection \ref{sus:nc-calculus}, and
  show that the model around the solution satisfying \eqref{deg}
  indeed looks like noncommutative $\U(1)\times G$
  Yang--Mills--Higgs model.
\end{exercise}
\begin{exercise}
 Show, that algebra \eqref{F-alg} can be a central extended Lie
 algebra. What does the model look like in this case? Try other
 closed algebras.
\end{exercise}
\begin{exercise}
  Generalize the map \eqref{map-2} to $\mathcal{G}$ an arbitrary $\su(n)$
  algebra.
\end{exercise}
\subsection{Solution with $\theta=0$}
A particularly interesting case is given by solutions with
$F_{\mu\nu}=\theta^{-1}_{\mu\nu}$ and $\phi_a=\const$. The
solution is highly degenerate in this case and a modification of
the analysis is needed. The action vanishes on such solutions,
therefore they correspond to absolute minima or vacua of the
model. Another property is that such solution also exist in finite
dimensional Hilbert spaces which is not the case of solutions with
nonzero $\theta$.

As one can see, equation \eqref{FF'-transform} is singular in this
limit and does not apply. Therefore a more detailed analysis is
needed.

In this case $\p'_\mu=\p_\mu+\mathbf{A}^{(0)}_\mu$, form a
commutative set,
\begin{equation}\label{com-p}
  [\p'_\mu,\p'_\nu]=0.
\end{equation}

Obviously, commuting operators fail to form an irreducible set,
unless the algebra is commutative which is not the case. However
let us still assume that the still form a complete set of
commutative operators, that is any operator commuting with all
$\p'_\mu$ can be expressed as a function of $\p_\mu$. For this
function to be unique, it is natural to require also all $\p_\mu$
to be functionally independent.\footnote{In the case when one or
several $\p'_\mu$ are functionally dependent on other operators
they can be dropped, coming to a smaller number of dimensions.} In
other words, this means that all $\p'_\mu$ are chosen so that they
are diagonalizable and any their common eigenvector $\ket{\xi}$ is
defined uniquely by its eigenvalues $\xi_\mu$,
\begin{equation}\label{eig-p}
  \p'_\mu \ket{\xi}=\xi_\mu\ket{\xi}.
\end{equation}

By a redefinition $\tilde{\p}_\mu=\tilde{\p}_\mu(\p')$ one can
make eigenvalues to be distributed uniformly in their range. Let
us denote this set as $\spec{\tilde{\p}}$. (We suppress the tilde
in the notations.)

In what follows, consider the case when $\spec{\p}$ coincides with
$\mathbb{R}^p$, i.e. eigenvalues $\xi_\mu$ are uniformly and
continuously distributed in the range from $-\infty$ to $+\infty$.
Then, one can introduce operators $\q^\mu$ defined as follows,
\begin{equation}\label{q-mu}
  \q^\mu\int\dd^p\xi\,f(\xi)\ket{\xi}= \ii\int\dd^p \xi \frac{\pd
  f(\xi)}{\pd\xi_\mu}\ket{\xi},
\end{equation}
where $\psi=\int\dd^p\xi\, f(\xi)\ket{\xi}$ is an arbitrary vector
of the Hilbert space written in the basis of
$\p_\mu$-eigenvectors.

It is not difficult to verify that $\p_\mu$ together with $\q^\mu$
form a $2p$-dimensional Heisenberg algebra (i.e. the Heisenberg
algebra of a $p$-dimensional particle),
\begin{equation}\label{2p-Heis}
  [\p_\mu,\q^\nu]=-\ii \delta_\mu^\nu.
\end{equation}

Let us note, that $\p_\mu$ and $\q^\mu$ together already form an
irreducible set of operators. Therefore, from this point on one
can apply the machinery of the section \ref{section:nc-space}.
After a computation one has that the action around the background
given by the solution \eqref{com-p} takes the form,\footnote{For
details we refer the reader to Ref.\cite{Sochichiu:2000ud}.}
\begin{equation}\label{com-act}
  S_{\mathrm{\theta=0}}=\int\dd^p p\dd^px\,\left(
  -\frac{1}{4g^2}F_{\mu\nu}(x,p)^2+\frac{1}{2}(\nabla_\mu
  \phi_a(x,p))^2-V_*(\phi(x,p))\right),
\end{equation}
where,
\begin{subequations}
\begin{align}
  &F_{\mu\nu}(x,p)=\pd_\mu A_\nu(x,p)-\pd_\nu
  A_\mu(x,p)+[A_\mu,A_\nu]_*(x,p),\\
  &\nabla_\mu\phi_a(x,p)=\pd_\mu\phi_a(x,p)+[A_\mu,\phi_a]_*(x,p),\\
  &[A,B]_*(x,p)\equiv A*B(x,p)-B*A(x,p),\\
  &A*B(x,p)=
  \left.\e^{\frac{\ii}{2}\left(\frac{\pd^2}{\pd x^\mu\pd p'_\mu}-
  \frac{\pd^2}{\pd x'{}^{\mu}\pd
  p_\mu}\right)}A(x,p)B(x',p')\right|_{\ov{x'=x}{p'=p}}.
\end{align}
\end{subequations}
The fields $A_\mu(x,p)$ and $\phi(x,p)$ are expressed in terms of
are expressed in terms of old one as follows,
\begin{subequations}
\begin{align}
  &(p_\mu+A_\mu(x,p))=\int\dd^px_{\mathrm{old}}K(x,p;x_{\mathrm{old}})
  (A^{(\mathrm{old})}_\mu(x_{\mathrm{old}})-\theta^{-1}_{\mu\nu}x_{\mathrm{old}}^\mu),\\
  &\phi_a(x,p)=\int\dd^px_{\mathrm{old}}K(x,p;x_{\mathrm{old}})
  \phi_a(x_{\mathrm{old}}),\\
  &K(x,p;x_{\mathrm{old}})=\tr[\mathbf{D}(x,p)\mathbf{D}_{\mathrm{old}}
  (x_{\mathrm{old}})],\\
  &\mathbf{D}(x,p)=\int\frac{\dd^pk\dd^pz}{(2\pi)^{2p}}
  \e^{\ii z(\p-p)+\ii k(\q-x)}
\end{align}
\end{subequations}

As we see, the noncommutative Yang--Mills--Higgs model turns to be
equivalent to a commutative Yang--mills--Higgs model with infinite
dimensional gauge group of diffeomorphisms.

\newpage
\section{Discussions and Outlook}
We discussed field theory on noncommutative spaces. In the Conne's
approach the noncommutative space is defined by the algebra of
continuous functions on it, its representation and a derivative
operator defined together with this algebra. The commutative space
appears to be a degenerate case of the above. The difference is
that the algebra defining the commutative space is Abelian and
this implies that there is the only representation of it which is
one-dimensional.

This is in contrast to what one has in the noncommutative case.
Noncommutativity of the space leads to existence of a class of
unitary equivalent representations instead of one single
representation. The physics, however, should look the same
irrespective to the chosen element of the equivalence class. Thus
we come to the notion of gauge invariance which is an intrinsic
feature of the noncommutative space. Therefore, the derivative
operator defining the noncommutative space can be identified with
the gauge field on it.

Thus, different gauge field configurations represent different
noncommutative spaces. Among these the most interesting ones are,
of course, those for which the gauge fields satisfy equations of
motion. The last are flat noncommutative spaces. Flat spaces with
unambiguous connection correspond to noncommutative spaces with
(flat) coordinates satisfying the Heisenberg algebra. The Hilbert
space representing it is infinite dimensional. Dropping out the
requirement of unambiguity allows one to have noncommutative
analogue of torus. Depending on defining parameters the
noncommutative torus can fit into a finite-dimensional subspace of
the Hilbert space. The study of the field theory on such spaces
remained so far beyond the scope of the present paper, although
they seem to be important at least as a regularization suitable
for numeric analysis.

Arbitrary ``deformations'' of noncommutative space including ones
changing the metric and topology as defined by the Connes geometry
are encoded in gauge fields. In this sense they play the also role
of gravity. If it is so, it would be interesting to separate the
gravity component of the noncommutative gauge theory.

So far we considered everything in purely classical approach. The
quantum theory even in the perturbative approach is known to face
some problems with renormalizability due to so called IR/UV mixing
\cite{Minwalla:1999px,VanRaamsdonk:2000rr}. Taking this into
account some of the above results can be generalized to the
quantum level \cite{now}.

\subsection*{Acknowledgements.} Different parts of this work were
reported at various seminars and Conferences in BLTP, Steklov
Mathematical Institute, Crete University, L\"udwig-Maximillian
University of Munich, etc. I benefited from discussions with many
people. In particular it is a pleasure to acknowledge ones with,
K.~Anagnostopoulos, I.~Aref'eva, P.~Aschieri, G.W.~Gibbons,
J.~Harvey, A.~Isaev, B.~Jur\v{c}o, E.~Kiritsis, A.~Nersessian,
A.~Slavnov, J.~Wess.
\newpage
\appendix
\section{Seiberg--Witten Map}\label{app:swm}
In this appendix we give a brief review on the Seiberg--Witten map
\cite{Seiberg:1999vs}. The bibliography on Seiberg--Witten map and
its applications is vast
\cite{Liu:2000mj,Garousi:2000ci,Bytsko:2000di,Jurco:2000dx,Jurco:2001kp,%
Bichl:2001nf,Garousi:2001jj,Brace:2001fj,Wyllard:2001ye,%
Bichl:2001yf,Kraus:2001xt,Picariello:2001mu,Banerjee:2001un,Suo:2001ih,%
Fidanza:2001qm}.

The Seiberg--Witten map was proposed as a map realizing
equivalences in low energies effective models for superstring
theory. The effective model in the presence of the constant
background field $B_{\mu\nu}$ computed in zero slope limit using
two different regularization schemes, namely in Pauli--Villars and
split-point regularization respectively, appears to be different
since in the first case it is a commutative theory with a
background constant field $B_{\mu\nu}$ while in the last one it is
a noncommutative model with no background field $B_{\mu\nu}$, but
whose noncommutative parameter $\theta^{\mu\nu}$ approaches the
value $(B^{-1})^{\mu\nu}$. The consistency requires this effective
models to be equivalent. As it was proposed in
\cite{Seiberg:1999vs}, this equivalence can be realized by a map
which relates the field configurations in these two cases. Beyond
this one can also consider e regularization scheme which is
intermediate between these two cases, and therefore the
equivalence should extend to arbitrary noncommutativity
parameters.

Since this is a map of gauge models, the natural requirement is
that gauge equivalent configurations should map to gauge
equivalent. Let $A_\mu(x)$ be the commutative gauge field and
$A'_\mu[A](x)$ or shortly $A'[A]$ be its image in the
noncommutative model, and let $U=\e^{\ii\lambda(x)}$ be an Abelian
gauge transformation,
\begin{equation}\label{ab-gauge-tr}
  A^U_\mu=A_\mu+\pd_\mu\lambda.
\end{equation}
Then there exists a noncommutative gauge transformation $U'[A,U]$,
which leads to the image of $A^U_\mu$,
\begin{equation}\label{SW-eq}
  A'[A^U]=(A')^{U'[A,U]}[A]\equiv
  U'{}^{-1}*A'[A]*U'+U'{}^{-1}*\dd U',
\end{equation}
where $U'[A,U]=\e^{\ii \lambda'[A,U=\e^{\ii\lambda}]}$, and in the
last expression we suppressed the functional arguments of $A'$ and
$U'$. The equation \eqref{SW-eq} is called \emph{Seiberg--Witten
equation}, and, respectively the map satisfying it is called
\emph{Seiberg--Witten map}.

For a small variation $\delta \theta^{\mu\nu}$ of the
noncommutativity parameter Seiberg--Witten equation \eqref{SW-eq}
takes the following infinitesimal form,
\begin{multline}\label{SW-inf}
  \delta A'_\mu(A^{(1+\lambda)})-\delta A'_\mu(A)-\pd_\mu \delta\lambda'
  -\ii[\delta\lambda',A_\mu]_*
  -\ii[\lambda,\delta A'_\mu]_* =\\
  -\frac{1}{2}\delta\theta^{\alpha\beta}
  (\pd_\alpha\lambda*\pd_\beta A_\mu+\pd_\beta
  A_\mu*\pd_\alpha\lambda),
\end{multline}
where $\delta A'_\mu(A)$ and $\delta\lambda'$ are infinitesimal
maps of the gauge field and gauge parameter respectively,
\begin{subequations}
\begin{align}
  &A'_\mu=A_\mu+\delta A'_\mu[A],\\
  &\lambda'=\lambda+\delta\lambda'[A,\lambda].
\end{align}
\end{subequations}

In \cite{Seiberg:1999vs} the following solution to \eqref{SW-inf}
was proposed,
\begin{subequations}\label{SW-map}
\begin{align}
  &\delta
  A_\mu=-\frac{1}{4}\delta\theta^{\alpha\beta}[A_\alpha*(\pd_\beta
  A_\mu+F_{\beta\mu})+(\pd_\beta A_\mu+F_{\beta\mu})*A_\alpha],\\
  &\delta\lambda=\frac{1}{4}\delta\theta^{\alpha\beta}
  (\pd_\alpha\lambda*A_\beta+A_\beta*\pd_\alpha\lambda).
\end{align}
\end{subequations}

This solution, however, is by far not unique. For example, one can
make a gauge transformation of either commutative field,
\begin{subequations}\label{SW-gt}
\begin{equation}
  A_\mu\to A^g_\mu,\qquad \e^{\ii\lambda}\to g^{-1}\e^{\ii\lambda},
\end{equation}
or,
\begin{equation}
  A'_\mu\to (A')^{g'}_\mu,\qquad U'\to (g')^{-1}U'.
\end{equation}
\end{subequations}
the transformed quantities will continue to satisfy \eqref{SW-inf}
or \eqref{SW-eq}. In particular $g$ can depend on $A$ and
$\lambda$, and respectively $g'$ on $A'$ and $\lambda'$, in this
case expressions \eqref{SW-map} change considerably.

\newpage
\section{K-theory and Morita Equivalence}\label{app:kt}
This appendix contains a short introduction to K-theory and Morita
equivalence. K-theory and Morita equivalence relation to string
theory and noncommutative geometry is discussed in the following
papers,
\cite{Horava:1998jy,Schwarz:1998qj,Gukov:1999qb,Diaconescu:2000wz,%
Diaconescu:2000wy,Witten:2000cn}. (For a review of K-theory see
\cite{Brodzki:1996}.)

Consider a $C^*$-algebra $\A$, or an associative complex algebra
with involution ``$^*$''. We will mainly think about the algebra
of complex functions on a noncommutative space. (In this case it
is a noncommutative algebra.) Let $E$ be its left module i.e.,
\begin{equation}\label{left_m}
  a(m)=am\in E,\qquad (a'a)(m)=a'(am)=a'am,
\end{equation}
for arbitrary $m\in E$, and $a,a'\in \A$. Right module is defined
in a similar way but with consequent action of elements of $\A$
from the right.

The algebra $\A$ itself and its tensor products
$\A\otimes\A\otimes\dots\otimes\A$ is a primitive example of both
left and right modules, such modules are called \emph{free}. A
module $E$ for which exists another module $E'$ such that $E\oplus
E'$ is free is a \emph{projective} one. (It is clear that $E'$ is
also a projective module.) The set of left or right projective
modules form a semigroup with respect to the direct sum operation.
This semigroup can be ``upgraded'' to a group as follows.

Consider pairs of modules $(E,F)$, with the composition rule
$(E,F)+(E',F')=(E\oplus E',F'\oplus F)$ and the equivalence
relation $(E,F)\sim (E\oplus G,F\oplus G)$, for arbitrary module
$G$. This equivalence classes form a group whose unity is given by
$(G,G)$-pairs and the opposite element to $(E,F)$ is given by
$(F,E)$,
\begin{equation*}
  (E,F)+(F,E)=(E\oplus F,E\oplus F)\sim (G,G).
\end{equation*}
This trick is similar to one used to extend the set of positive
numbers to real ones. The group one gets in such a way is called
the K$(\A)$ or if $\A$ is the algebra of functions on some space
$\M$ it is denoted as K$(\M)$.

Let us equip our left or right projective module $E$, with an
$\A$-valued product $\pr{\ ,}$, satisfying,
\begin{align}\label{prod1}
  &\pr{m,m'}^*=\pr{m',m} \\ \label{prod2}
  &\pr{am,m'}=a\pr{m,m'} \\ \label{prod3}
  &\pr{m,m'}\text{ is a positive element in }\A.
\end{align}
In other words, if we assume that the algebra $\A$ is equipped
with trace, $\tr a^*=(\tr a)^*$, then the eqs.
(\ref{prod1}-\ref{prod3}) mean that $\tr \pr{m,m'}$ should define
a nondegenerate scalar product whose adjoint is compatible with
the involution. The interesting most case of the \emph{full}
module $E$ is when the linear span of the range of $\pr{\ ,}$ is
dense in $\A$.

One can introduce connection $\na_\alpha$ on the $\A$-module $E$
with respect to infinitesimal automorphisms of $\A$: $a\to
a+\delta_\alpha a$, labelled by some element $\alpha$, which
satisfies,
\begin{equation}\label{connection}
  \na_\alpha(am)=a\na_\alpha(m)+\delta_\alpha a m,
\end{equation}
and it is linear in $\alpha$. Using this connection one can built
the curvature associated to it,
\begin{equation}\label{curvature}
  F_{\alpha\beta}=[\na_\alpha,\na_\beta]-\na_{[\alpha,\beta]}.
\end{equation}

$\A$-linear maps $T:E\to E$ which have an adjoint with respect to
the product (\ref{prod1}-\ref{prod3}) and commute with the action
of $\A$ on $E$ form the algebra $\End E$ of endomorphisms of the
$\A$-module $E$.

By definition an algebra $\B$ is Morita equivalent to $\A$ if it
is isomorphic to $\End E$ for some complete module $E$.

There exists the following criterium for Morita equivalence of two
algebras $\A$ and $\B$. A left $\A$-module $P$ which is also a
right $\B$-module is called $(\A,\B)$-bimodule. Assume that $P$ as
$\A$- and $\B$-module is equipped with $\A$-valued product $\pr{\
,}$, and $\B$-valued product $\prB{\ ,}$, and it is full as both
$\A$- and $\B$-module. When it exists such a module is called
$(\A,\B)$ \emph{equivalence bimodule}, in this case algebras $\A$
and $\B$ are Morita equivalent. The Morita equivalence allows one
to establish relations between various structures of the
equivalent algebras and their modules, like endomorphisms,
connections, etc.

It is conjectured
\cite{Horava:1998jy,Schwarz:1998qj,Diaconescu:2000wz,Diaconescu:2000wy,%
Witten:2000cn}, that Morita equivalent algebras in string theory
correspond to physically equivalent systems e.g. related by
duality transformations. In noncommutative theory the gauge models
on the dual tori are also known to be Morita equivalent
\cite{Connes:1998cr,Schwarz:1998qj}.

\newpage
 \bibliographystyle{JHEP}
 \bibliography{noncom}

\end{document}